%% file: writeup.tex
\documentclass[12pt]{article}
\usepackage{cite,colordvi,epsf,amssymb,color}
\input{epsf2}

\newsavebox{\fmbox}

\newcommand{\uxm}{U_{x,\mu}}
\newcommand{\be}{\begin{equation}}
\newcommand{\ee}{\end{equation}}
\newcommand{\ba}{\begin{eqnarray}}
\newcommand{\ea}{\end{eqnarray}}

\newcommand{\Tr}{\ensuremath{\mathop{\rm Tr}}}

\newcommand{\gsim}{\ensuremath{\mathrel
               {\raise2pt\hbox to 8pt{\raise -5pt\hbox{$\sim$}\hss{$>$}}}}}
\let\orighspace\hspace
\renewcommand{\hspace}{\vrule width0pt\relax\orighspace}

\newcommand{\lqcd}{\Lambda_{QCD}}
\begin{document}
\begin{center}
\parbox{\hsize}{\vspace{-5em}
\hbox to \hsize
{\hss \normalsize HU-EP-04/10}}

{\Large{Effective Field Theories and Quantum Chromodynamics on the Lattice}}

A.~Ali~Khan \\
{\it Institut f\"ur Physik, Humboldt-Universit\"at 
zu Berlin,  \\ 12489 Berlin, Germany}
\end{center}

\begin{abstract}
We give a selection of results on spectrum and decay constants
of light and heavy-light hadrons. 
Effective fields theories relevant for their lattice calculation,
 namely non-relativistic QCD (NRQCD) for heavy quarks on  the lattice and  Chiral Perturbation Theory for light 
quarks, are briefly discussed.
\end{abstract}
\section{INTRODUCTION}
The Standard Model of the strong and electroweak interactions is based 
on a $SU(3)\times SU(2)\times U(1)$ gauge symmetry with three generations of quarks and leptons as fermionic
matter fields and a scalar field, the Higgs, which is responsible for the masses of the weak $SU(2)$ gauge
bosons and of the fermions. For a recent review about the status of the Standard Model and new physics
see e.g.~\cite{wilczek2003}.

The $SU(3)$ `sector' of the model is Quantum Chromodynamics (QCD), a gauge theory of the 
strong interaction.
With relativistic Dirac quarks, the model can be described classically by the Lagrangian
\ba
{\cal L}_{QCD} &=& \overline{q}\left(i\gamma_\mu D^\mu -m\right) q 
 -\frac{1}{4}  F^c_{\mu\nu}F^{c\mu\nu}. \label{eq:qcd}
\ea
The $q$ fields are 4-component Dirac spinors, and the $D_\mu$ are covariant derivatives,
e.g. $D_\mu \equiv \partial_\mu - ig_s A_\mu^c t^c$ with  $[D_\mu,D_\nu] = -ig_sF^c_{\mu\nu}t^c$,
where $F^c_{\mu\nu}$ are the field strength tensors, $g_s$ is the coupling constant,   and 
the $t^c$ are generators of $SU(3)$ in the fundamental representation. 
A consequence of the self-interactions among the  gluon fields $A_\mu^c$ is
asymptotic freedom, i.e. the interactions between particles become weak at short distances and  
can be described with 
perturbation theory in the strong coupling  $\alpha_s = g_s^2/(4\pi)$. At larger distance, 
the forces become strong, and non-perturbative methods are necessary to 
understand how hadron masses arise and whether it is
possible to explain the hadron spectrum from first principles within the theory of 
strong interactions. 

Although rather successful, the Standard Model by itself does not seem completely satisfactory.
On the experimental side, recent discoveries 
such as neutrino mixings, new results from accelerator experiments~\cite{belle2003,g-22004}
and indications for `dark energy' in the cosmos indicate a need for an extension of the model.
The Higgs particle has not yet been found; recent reviews of the status of Higgs searches
are~\cite{djouadi2005}.  
Further there are theoretical motivations to search for 
physics beyond the Standard Model (for a discussion see e.g.~\cite{wilczek1998}).  
The Standard Model contains a considerably 
large set of coupling constants and masses as input parameters. It does not explain 
the values of typical energy scales such as the masses of the weak gauge bosons.

A strategy in the research is to simultaneously measure as many physical quantities as possible, test
the results for self-consistency within the Standard Model and search for indications of new
physics.
Among the most interesting search grounds are the elements of the Cabibbo-Kobayashi-Maskawa (CKM) matrix 
which parameterizes the flavor changing weak currents and provides a mechanism for CP violation within the 
Standard Model. Those CKM matrix elements which are relevant to reactions of heavy, for example $b$ and $c$, 
quarks are at present studied intensively in experiment and theory. 
We introduce the CKM matrix with an emphasis on $B$ meson decays
in the framework of the weak effective theory in Section~\ref{sec:CKM}.  The status of the 
CKM matrix is reviewed in~\cite{ali2003}.
For a review about recent results on quark masses see Ref.~\cite{gupta2003}.

Description of the long-range interactions of QCD requires non-perturbative techniques.
Using a four-dimensional lattice description of space and time it is possible to 
calculate matrix elements numerically on a computer within a path integral formalism.
A brief introduction to the lattice formalism is given in Section~\ref{sec:lattice}; for detailed recent
reviews see~\cite{latreviews}.

Ideally, the lattice extent $L$ should be much larger than the extent or the  Compton
wavelength of the particles that are supposed to be described, and the inverse lattice spacing $a$
should be much larger than the masses and momenta in the theory in order to avoid cutoff effects.
The lightest hadrons, the pions, have a mass of around 140 MeV, whereas the $B$ meson has a mass
of $5.28$ GeV and contains a heavy quark with a mass of 5 GeV. The problem is how lattice simulations can 
accommodate this large range of scales.

\begin{center}
{\small
\begin{tabular}{cccccc}
Ideally  
         &$L^{-1}$& $\ll$ &          masses and  & $\ll$ & $a^{-1}$  \\
         &     &       &  energy splittings  &       &        \\
         &     &       &     &       &        \\
\hline
In Reality &            &       &                       &       &              \\
           & $L=2-3$ fm &       &                       &       & $a^{-1}=2-4$ GeV    \\
           & ($L^{-1}=0.07-0.1$ GeV) &                     &        &                   \\
           & finite size effects&   &                    &       & cutoff effects       \\
\end{tabular}
}
\end{center}

To calculate  properties of hadrons with $b$ quarks on the lattice,  one can for example simulate at lighter quark 
masses where discretization errors are under better control,  and use extrapolations in the heavy quark mass.
Fortunately the energy level splittings of $b$ hadrons are much smaller than their masses:
of the order of $\lqcd = 200-500$ MeV or smaller, where $\lqcd$ is the energy scale where QCD becomes 
non-perturbative. 
The dynamics of the heavy quarks can be accounted for as small corrections proportional to powers of the
inverse heavy quark mass.
This is the basis for effective field theories developed for heavy quarks: Heavy Quark Effective Theory 
(HQET)~\cite{isgur1992}
and non-relativistic QCD (NRQCD)~\cite{thacker1991,lepage1992} (for reviews see~\cite{neubert1994}),
which can be
used to simulate heavy quarks directly on the lattice while avoiding large discretization errors due to the
large mass. We discuss NRQCD in Section~\ref{sec:effective}.

Practical simulations with light quarks are computationally expensive and sensitive to the finite lattice
volume. Therefore one often uses quark masses much heavier than $u$ and $d$ quark 
masses  and extrapolates the results to the physical values of the quark masses. 
A formalism for this can be derived using chiral perturbation theory ($\chi PT$), an expansion around the chiral
(zero quark mass) limit describing low-energy degrees of freedom of QCD such as pions and nucleons. This is introduced 
in Section~\ref{sec:chiral}.

Lattice results for light and heavy-light hadron masses
and heavy-light current matrix elements are discussed in Section~\ref{sec:results}. 
\subsection{Heavy quark decays\label{sec:CKM}}
\begin{figure}[thb]
\begin{center}
\vspace{-1cm}
\centerline{
\hspace{2.8cm}
\epsfysize=2.8cm \epsfbox[300 50 500 400]{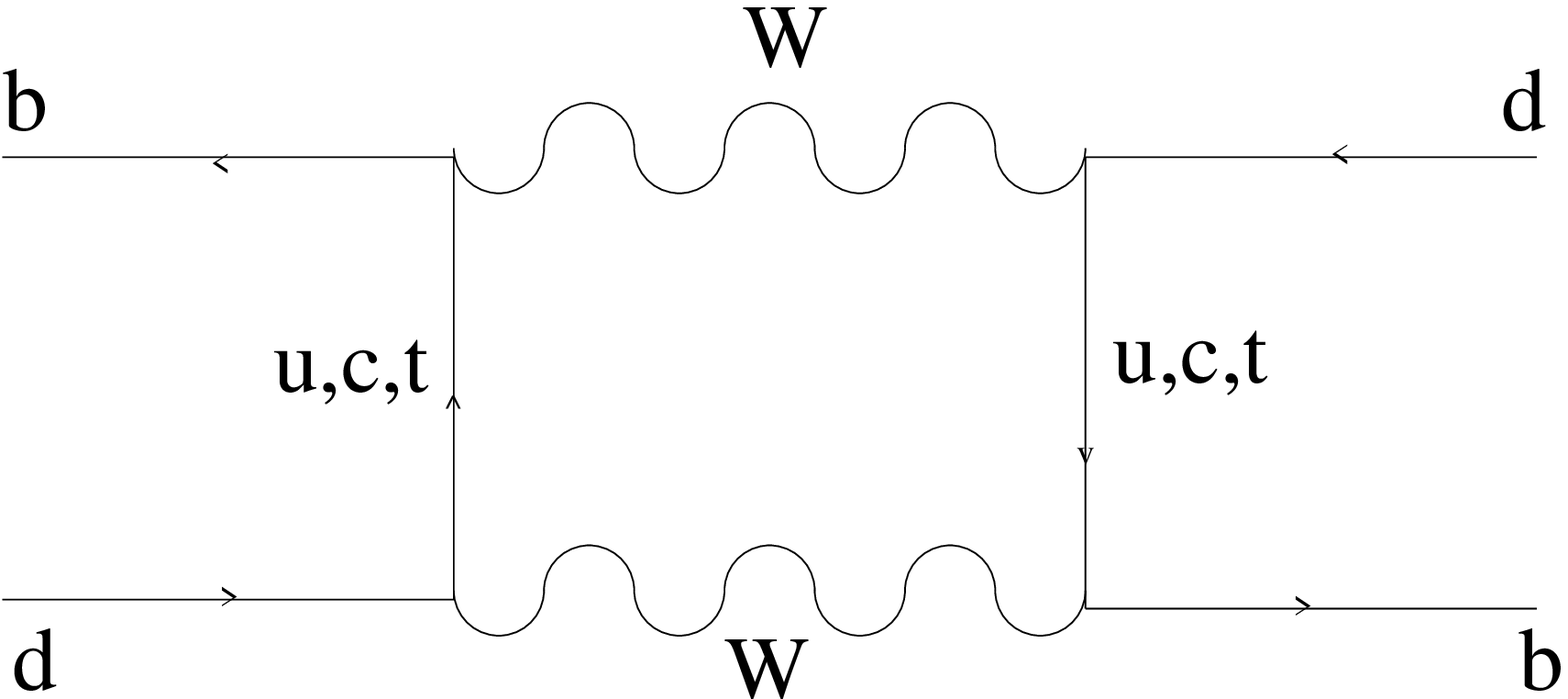}
\hspace{1cm}
\epsfysize=2.8cm \epsfbox[50 50 220 400]{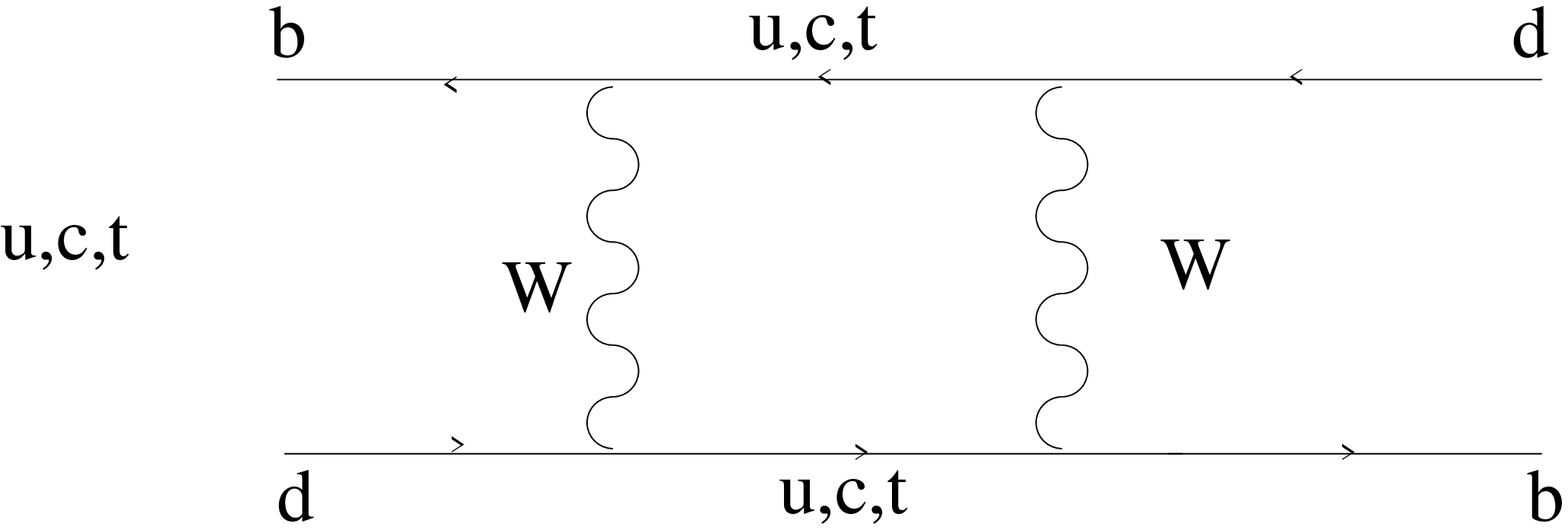}
\hspace{4.5cm}
\epsfysize=2.8cm \epsfbox[0 20 400 270]{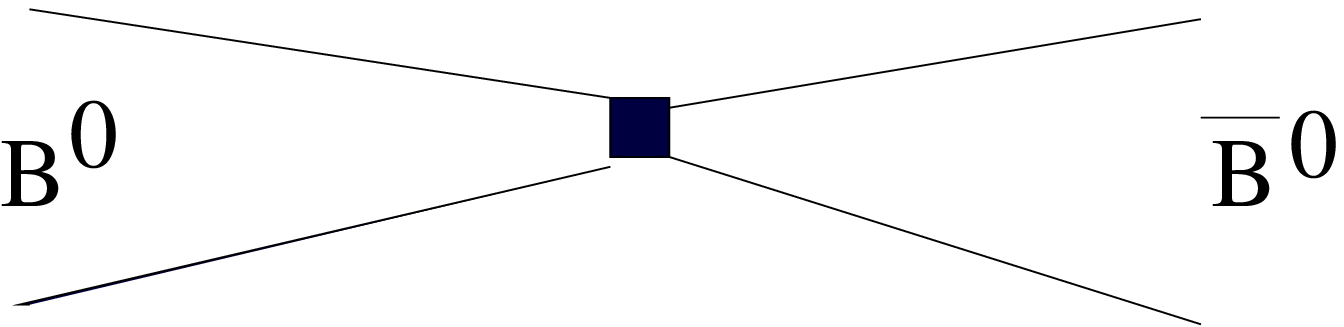}
}
\end{center}
\caption{Box diagrams describing $B^0-\overline{B^0}$ mixing in the electroweak theory (left and middle) and the 
weak effective theory (right).}
\label{fig:box}
\end{figure}

Study of weak decays of quarks is of interest for determinations of elements of the CKM matrix which 
parameterizes the mixings of quark generations in the Standard Model:
\be
{\cal L} = -\frac{g_2}{\sqrt{2}} \left(\overline{u}_L\overline{c}_L\overline{t}_L\right)
\gamma^\mu (V_{q_1q_2}) \left(
\begin{array}{c}
d_L \\
s_L \\
b_L \\
\end{array}
\right) W^\dagger_\mu + \mathrm{h.c.}\;,
\ee
where $u_L,d_L,c_L,s_L,t_L,b_L$ are left-handed quark spinors, $W_\mu$ a charged weak gauge boson 
and $g_2$ the weak gauge coupling.
$V$ is a unitary matrix. There are indications that some of its elements have a non-trivial
complex phase giving rise to CP violation. 
The CKM matrix elements with presently the largest uncertainties are the ones relevant to decays or mixings 
of the $b$ quark: $V_{cb}$, $V_{ub}$ and $V_{td}$. 
$|V_{ub}|^2$ describes for example the leptonic meson decay $B^+ 
\rightarrow l^+ \nu_l$,
where $l$ is a lepton ($e$, $\mu$ or $\tau$) and $\nu_l$ the corresponding neutrino and
semileptonic decays into a light meson ($B \rightarrow \pi, \rho, \omega$) and a lepton-neutrino pair, and
$|V_{cb}|^2$ determines the semileptonic decay $B \rightarrow D$
and a lepton-neutrino pair.
$|V_{td}|^2$ is proportional to the oscillation frequency between
the mass eigenstates of the $B^0-\overline{B^0}$ mixing, which is described by the 
left and middle diagrams in Figure~\ref{fig:box} in the electroweak theory.
Processes at energy scales much less than the $W$ boson mass can be calculated within
the weak effective theory where interactions mediated by the $W$ or $Z$ particles can be
described  by point interactions. $B^0-\overline{B^0}$ is described by the
third diagram in Fig.~\ref{fig:box}.

To relate the weak processes between quarks with exclusive reaction rates of mesons, one uses 
form factors which get  contributions from long-distance QCD interactions,
and therefore have to be calculated nonperturbatively. This can be done from first principles
using the lattice.
In the effective theory the $B$ meson decay is described by a matrix element of the
heavy-light axial vector current
\be
\langle 0|A_\mu(x)|B(p) \rangle = if_B p_\mu e^{-ipx}, \label{eq:axial}
\ee
where $f_B$ is the $B$ decay constant.
The branching ratio for the decay $B^+ \rightarrow l^+\nu_l$  is 
\ba
BR(B^+ \rightarrow l^+ \nu_l) &=& \frac{G_F^2 m_B m_l^2}{8\pi}\left(1-
\frac{m_l^2}{m_B^2} \right)^2  f_B^2 |V_{ub}|^2 \tau_B, 
\ea
where $G_F= g_2^2/(8M_W^2)$ is the Fermi constant and $\tau_B$ the $B$ lifetime. If $f_B$ is known, 
$|V_{ub}|$ can in principle be determined experimentally from this decay. 
There exist only experimental upper bounds on $f_B$, $f_{B_s}$ and
$f_D$, but there are results on  $f_{D_s}$:
\ba
f_{D_s} &=& 280(17)(25)(35) \mbox{ MeV~\protect\cite{cleo_fDS} and} \nonumber \\
 &=& 285(19)(40) \mbox{ MeV~\protect~\cite{aleph_fDS}}. \nonumber \label{eq:fDs}
\ea 
In the weak effective theory, the form factor for the $B^0-\overline{B^0}$ mixing matrix element can 
be parameterized as $f_B^2B_B$, where the ``bag parameter'' $B_B$ quantifies to what extent the matrix 
element is described by $B$-to-vacuum currents:
\be
B_B = \frac{3}{2}\frac{\langle \overline{B^0} | (\overline{b}_L\gamma_\mu d_L)(\overline{b}_L\gamma_\mu d_L)|B^0
\rangle}{\langle \overline{B^0} |\overline{b}\gamma_\mu\gamma_5 d|0\rangle
\langle 0|\overline{b}\gamma_\mu\gamma_5 d|B^0\rangle }.
\ee
The oscillation frequency of the mass eigenstates is proportional to the mass difference and
related to the form factors by
\be
\Delta M_d \propto  |V_{tb}^\ast V_{td}|^2 f_B^2B_B.
\ee

\section{QCD ON THE LATTICE\label{sec:lattice}}
\subsection{Gauge fields}
Matter fields, e.g.\ quarks, sit on the lattice sites which are separated by a
spacing $a$. The gauge fields on the lattice are represented by fields $U_{x,\mu}
= \exp[ig_s a \int_{x+\mu}^xdzA^c_\mu(z)t^c]= \exp[-ig_s a A^c_\mu(x)t^c]$,
parallel transporters between neighboring lattice sites.
Line integrals of gauge fields over closed paths are called Wilson loops. The smallest 
($1\times1$) Wilson loop on the lattice is a product of gauge links over the the nearest neighbors, 
called plaquette, 
$P^\dagger_{\mu\nu} = \uxm U_{x+\mu,\nu}U^\dagger_{x+\nu,\mu}U^\dagger_{x,\nu}$:

\begin{center}
\epsfysize=2cm \epsfbox{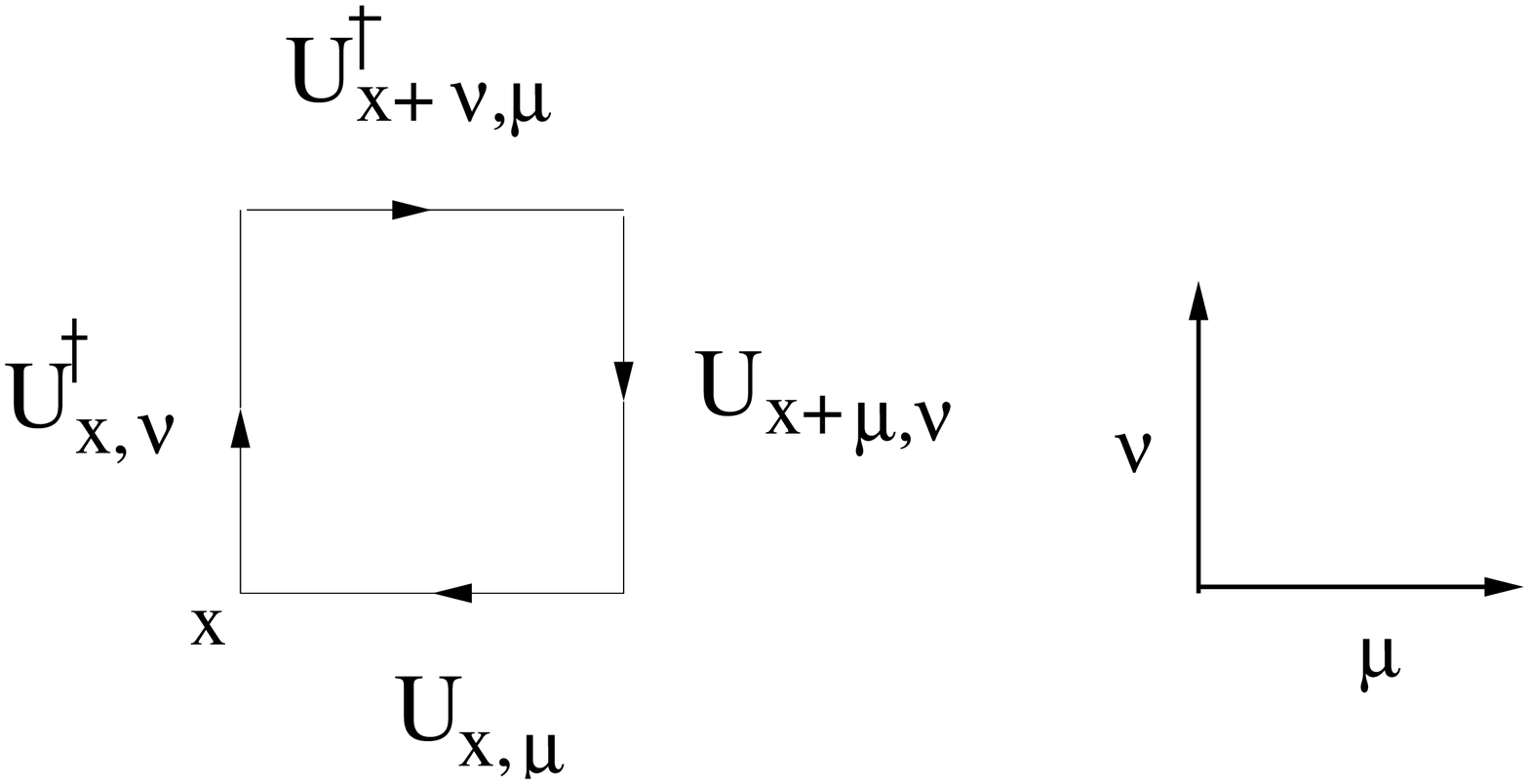}
\end{center}

By expanding  around $a \simeq 0$, one finds a tree-level
relation between the plaquette and the continuum field strengths. 
Thereby one obtains the Wilson (or `plaquette') lattice action as a discretization of the continuum gauge
field action of Eq.~(\ref{eq:qcd}),
\be
S_g = \frac{-\beta}{N_c}\sum_\Box Re\Tr\Box, \,\, 
\beta = \frac{2N_c}{g_s^2}, \mbox{ $N_c$: number of colors},
\ee
which has lattice spacing errors at $O(a^2)$. To further reduce discretization effects, actions
can be improved. For gauge field actions, this consists of adding larger Wilson 
loops ($1\times2$, ...). Improvement can be done  by removing discretization effects 
order by order in $a$ (originally suggested by Symanzik for scalar field theory~\cite{symanzik1980}, and
developed into an improvement program for on-shell quantities in QCD by Ref.~\cite{luscher1985}), or with 
renormalization group methods to obtain renormalization group improved (RG)~\cite{iwasaki1985} or 
perfect~\cite{hasenfratz1994} actions.

At typical values of $\beta$ in lattice simulations,
there are large  corrections due to gauge field loops on the lattice which shift the expectation
value of $U$ substantially with respect to the free field value, one. 
The perturbative corrections can be reduced with `mean-field' (or `tadpole') improvement~\cite{lepage1992}: 
the gauge links $U$ are divided by their expectation value which can 
be calculated in perturbation theory or determined nonperturbatively in simulations.
\subsection{Lattice fermions\label{sec:ferm}}
Discretization of the Euclidean continuum Dirac action by substituting the covariant
derivatives by covariant symmetric lattice differences gives the 'naive' lattice fermion action
\ba
S_F& = &a^4\left( \frac{1}{2a} \sum_{x,\mu}\overline{q}_x\gamma_\mu\left[
\uxm q_{x+\mu} - U^\dagger_{x-\mu,\mu}q_{x-\mu}\right]
 +\sum_x m\overline{q}_x q_x\right),
\ea
which is chirally symmetric if $m\rightarrow 0$ and has $O(a^2)$ errors.
However, the naive discretization leads to a flavor multiplication, the so-called 
'doublers'. 
At $m=0$, the free fermion propagator has a pole at $k_\mu = 0$ as in the continuum but 
also poles at $k_\mu = \pi/a$.  There are 16 species of fermions, which occur in pairs of
opposite chirality.

Wilson's solution to the doubling problem is to add a term of the form $a^4\sum_x
a\overline{q}_x\Delta q_x$
to the action, where $\Delta$ is a covariant second derivative.
The doublers obtain masses which remain finite in lattice units: $ma \neq 0$ if $a\rightarrow 0$.
Chiral symmetry receives corrections at 
$a \neq 0$, and $O(am), O(ap)$ discretization errors occur. $O(a)$ errors can be removed from the
action with the clover term proportional to $a^4\sum_x\overline{q}_x\sigma_{\mu\nu}G^{\mu\nu}q_x$, where $G_{\mu\nu}$
is a discretized version of the field strength tensor using four neighboring plaquettes~\cite{SW1985}.
The coefficient of the clover term can  be calculated in
perturbation theory (a common choice is at tree-level using tadpole-improvement). 
Most recent calculations use a non-perturbative determination of the clover coefficient~\cite{cSW}
and are $O(a)$ improved to all orders in perturbation theory.

Staggered fermions are obtained from a spin-diagonalization of naive fermions:
\be
q_x = \gamma_x \chi_x \, , \;\; \overline{q}_x = \overline{\chi}_x \gamma_x^\dagger\;,
\ee
with $\gamma_x = \gamma_1^{x_1}\gamma_2^{x_2}\gamma_3^{x_3}\gamma_4^{x_4}$, where $x = 
(x_1,x_2,x_3,x_4)$. The $\chi$ fields are one-component. In the massless case, the theory has
a  $U(1)\times U(1)$ chiral symmetry at finite $a$ and a $U(4)\times U(4)$ chiral symmetry 
in the continuum limit. Discretization errors are $O(a^2)$. Improvement is possible
by adding higher dimensional operators.

Lattice formalisms for doubler-free, chiral fermions are  given in
\cite{domainwall,overlap,hasenfratz1998}.

\subsection{Extracting physical quantities}
Green functions can be calculated by  evaluating the path integral over the lattice degrees of
freedom numerically. For example, a
two-point function of a field $O(x)$ which can be composed of more elementary fields 
$\{\phi_i(x)\}$ is given
by
\be
\langle O^\dagger(x) O(0) \rangle = \frac{1}{Z}\int {\cal D}\phi O^\dagger(x)
O(0) e^{-S[\phi]}\;,
\ee
where ${\cal D\phi}$ denotes integration over all dynamical fields (gauge, fermion, etc...)
in the theory. To determine for example $f_B$ from the lattice, it is necessary to calculate the 
renormalization factors to match the unrenormalized lattice matrix element of the axial vector
current to the corresponding matrix element in continuum QCD. 

Ideally, these calculations are done at various values of the lattice spacings, and the
continuum estimate is obtained by extrapolating as a function of $a$ to
$a \rightarrow$ 0. In practice, some lattice calculations are performed only at one or two
values of $a$, in which case a continuum limit cannot be taken, and the discretization effects
have to be included into the estimate of systematic errors. With NRQCD calculations, higher dimensional 
operators are included as discussed in Section~\ref{sec:NRQCD}, and an $a \rightarrow 0$ extrapolation 
cannot be done out of principle. 
Calculation at several values of $a$ then serves to determine the systematic error from keeping
the lattice spacing finite.

In full QCD, the path integral includes
gauge and fermionic fields $ \int {\cal D}U{\cal D}\overline{q}{\cal D}q$.
To decrease computational expenses, many calculations are done in the 
quenched approximation, i.e. the vacuum polarization due to quark loops is neglected.

To use lattice results in phenomenology, it is necessary to estimate systematical errors 
as accurately as possible. The most important sources are:
\begin{itemize}
\item Finite lattice spacing 
\item Finiteness of lattice volume
\item Quenching (unphysical number of dynamical quarks)
\end{itemize}
\section{EFFECTIVE THEORIES AND THE LATTICE\label{sec:effective}}
\subsection{NRQCD\label{sec:NRQCD}}
Non-relativistic QCD (NRQCD) is an effective theory formulated for heavy quarks assuming that their 
dynamics is non-relativistic, with  correction terms which can be added within a systematic
expansion. For quarkonia the higher order interactions are arranged in a $v^2$
expansion, where $v$ is the heavy quark velocity (see e.g.~\cite{lepage1992}).
In heavy-light systems it is an expansion in $v$ or $1/M$, where $M$ is the heavy quark mass.
At infinite mass, the heavy quark is just a source of the color electromagnetic field, whereas
at finite $M$, there is a recoil of the heavy quark due to the interaction with soft gluons with
typical momenta of $O(\lqcd)$.
One can argue that the heavy  quark 
momentum $P_Q$ and light quark momentum $p_q$ are equal due to momentum conservation within the rest
frame of the meson:
\ba
M v \simeq P_Q &=& p_q \sim O(\Lambda_{QCD}). 
\ea
Therefore $v \sim \Lambda_{QCD}/M \sim 0.1$ in $B$ mesons and should be a reasonable expansion parameter
to specify corrections to the $M \rightarrow \infty$ (static) limit.
Contributions at $O(1/M)$ are the kinetic and the spin-colormagnetic energy of the heavy quark; at $O(1/M^2)$ 
a heavy quark spin-orbit interaction and a Darwin term are added.
The $O(1/M^2)$ Lagrangian for the heavy quark is given by
\be
{\cal L} = \psi^\dagger(D_t + H)\psi, \label{eq:NRQCD}
\ee
with heavy quark Pauli spinor $\psi$ and the Hamiltonian 
\ba
H &=& -\frac{\vec{D}^2}{2M} - \frac{g_sc_4}{M} 
\vec{\sigma}\cdot\vec{B} \nonumber \\
&+& \frac{ig_sc_2}{8M^2}(\vec{D}\cdot\vec{E}-\vec{E}\cdot
\vec{D})  - \frac{g_sc_3}{8M^2}\vec{\sigma}\cdot(\vec{D}\times
\vec{E}-\vec{E}\times\vec{D}) - \frac{c_1(\vec{D}^2)^2}{8M^3}. \label{eq:NRQCD_cont}
\ea
The last term is the first relativistic correction to the kinetic energy of the heavy quark, which is
usually included in calculations at $O(1/M^2)$. 
The coefficients of the various terms can be found with matching calculations to full QCD in the 
continuum.  
In the lattice calculations described in Section~\ref{sec:results}, they are set to their tree-level value one,
using mean field improved gauge links.

Discretizing the Lagrangian~(\ref{eq:NRQCD}) one can simulate $b$ quarks directly on the lattice,
since there are no $O((aM)^n)$ discretization errors.
Errors $O(a^2\vec{p}^2)$ and $O(aMv^2/2)$ arising from discretization of the spatial and temporal 
derivatives in the 
NRQCD Lagrangian can be corrected for by adding further terms to the Hamiltonian~(\ref{eq:NRQCD_cont}). 
If matrix elements of operators
are to be calculated in this formalism, the $1/M$ corrections to the operators have to be 
taken into account as well. Simulation of the $1/M$ corrections within
HQET on the lattice using nonperturbative renormalization is discussed in~\cite{heitger2004}.

Other methods to avoid large discretization errors used in the calculations 
discussed here are to
simulate heavy quarks around the charm and extrapolate
to the $b$, or to use a non-relativistic interpretation of a Wilson or clover 
heavy quark action called FNAL~\cite{elkhadra1997} in this article. 
Refs.~\cite{elkhadra1997} and~\cite{aoki2002} formulate on-shell improvement programs for heavy quarks 
by adding further operators to the Wilson or clover action. For recent reviews
see also~\cite{bernard2001a,kronfeld2003}. 

\subsection{Chiral perturbation theory\label{sec:chiral}}
The chiral symmetry of massless QCD can be understood as  
spontaneously broken. As a  consequence one would expect massless Goldstone
bosons in the spectrum. One can identify the physical pions with the Goldstone bosons if 
small quark mass terms are included in the QCD Lagrangian  resulting in a small explicit breaking of 
chiral symmetry and a finite but small pion mass $m_\pi$. 
Typical momenta of low-energy interactions of pions will be $O(m_\pi)$. 
Chiral perturbation theory can be formulated as an 
effective theory expanding in powers of the pion mass and external momenta $p =  O(m_\pi)$,
the so-called $p$ expansion~\cite{gasser1985}. It is used to describe physical interactions at
low energy scales, for example pion-nucleon scattering. It gives predictions for the expansion of
hadron masses around the zero quark mass limit, which can be  used in the analysis of lattice
calculations to extrapolate lattice hadron masses  simulated at larger quark 
mass to the physical light ($u,d$) quark mass.
The leading dependence at $O(p^2)$ is expected to be linear in the quark mass $m_q$ where $m_q \propto m_\pi^2$, 
with corrections  at $O(p^3)$ etc.
To facilitate inclusion of interactions with baryons within the $p$ expansion, Heavy Baryon Chiral Perturbation Theory  
($HB\chi PT$) was developed~\cite{jenkins1991}, which works in the limit of infinite baryon mass. 

The quark masses in present simulations are such that pions are not much lighter than the lattice nucleons,
and a formulation with relativistic nucleons seems more appropriate~\cite{relativistic,becher1999}. 

For the pions one can use a representation based on their nature as
Goldstone bosons of the broken $SU(2)\times SU(2)$ 
chiral symmetry of massless QCD with two flavors of $u$ and $d$ quarks with
\be
{\cal L}_\pi = \frac{1}{4}f_\pi^2\Tr\left[\partial_\mu U^\dagger
\partial^\mu U +\chi^\dagger U + \chi U^\dagger)\right].
\ee
$U = \exp(\frac{i}{f_\pi} \vec{\tau}\vec{\pi})$ transforms according to the
$(2,2)$ representation of $SU(2)\times SU(2)$.
$\chi = 2B_0 {\cal M}$, ${\cal M}$ is the quark mass matrix and $B_0$ is
proportional to the chiral condensate.
The $\vec{\pi}$ are the pion fields, and the $\tau_i$ are Pauli matrices.
$f_\pi$ is the pion decay constant. 

The nucleon Lagrangian at lowest order $(O(p^1))$ is
\be
{\cal L}^{(1)} =  \overline{\Psi}\left(i\gamma_\mu D^\mu - m_0\right)\Psi + 
\frac{1}{2} g_A\overline{\Psi} \gamma_\mu \gamma_5 u^\mu \Psi,
\ee
with $u^2 = U$, $D_\mu = \partial_\mu + \frac{1}{2}[u^\dagger,\partial_\mu u]$ and
$u_\mu = iu^\dagger \partial_\mu U u^\dagger$. $\Psi$ is the Dirac spinor of the nucleon,
$m_0$ the nucleon mass in the chiral limit, and
$g_A$ the nucleon axial vector coupling in the chiral limit. The $O(p^2)$ Lagrangian is
given by
\ba
{\cal L}^{(2)} &=&  c_1\mathrm{Tr}(\chi_+)\overline{\Psi}\Psi
+ \frac{c_2}{4m_0^2}\mathrm{Tr}( u_\mu u_\nu)\left( \overline{\Psi}
D^\mu D^\nu \Psi + h.c.\right) \nonumber \\
&+& \frac{c_3}{2}\mathrm{Tr}( u_\mu u^\mu ) \overline{\Psi}\Psi
-\frac{c_4}{4}\overline{\Psi} \gamma^\mu \gamma^\nu [u_\mu,u_\nu]\Psi,
\ea
with $\chi_+ = u^\dagger \chi u^\dagger + u \chi^\dagger u$.
$\mathrm{Tr}()$ refers to the trace over the flavor indices.

Chiral extrapolation of nucleon masses from the lattice and calculation of 
the effect of the lattice size on nucleon masses are discussed in Section~\ref{sec:FS}.

If the pions are light compared to the inverse lattice extent, but the lattice size 
is not too small such that $1/L  \sim \epsilon$, where $\epsilon$ is a small parameter,
one can study fluctuations around the zero modes within the
so-called $\epsilon$ expansion of $\chi PT$~\cite{epsilon1987}. 
\section{LATTICE RESULTS\label{sec:results}}
\subsection{Setting the scale}
Masses and decay constants coming out of a simulation are at first dimensionless numbers in units of 
the lattice spacing.
\begin{figure}[thb]
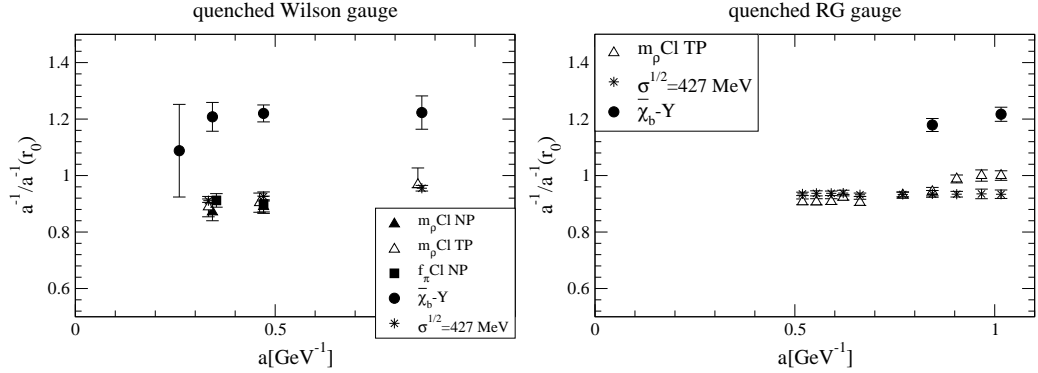

\begin{center}
\centerline{
\epsfysize=4.9cm \epsfbox{a_wilson.eps}
\epsfysize=4.9cm \epsfbox{a_rg.eps}
}
\end{center}
\vspace{-0.5cm}
\caption{Discrepancy of lattice spacings from various physical quantities 
on quenched lattices. Results are from 
\protect\cite{guagnelli1998,cppacs2002hadr,jlqcd2003hadr,davies1997,manke2000,hein2000,bowler2001}.
Average over spin orientations is denoted by an overbar.}
\label{fig:aquenched} 
\end{figure}
The value of the lattice spacing is  determined by calculating a suitable quantity $aM$ on the lattice and
adjusting the corresponding dimensionful quantity $M$ to its physical value. 
If the calculation is free of
systematic errors such as lattice spacing, finite volume and quenching effects, using any quantity should
give the same result. In practical calculations all of those errors can occur. Then, 'suitable' means
that systematic errors of the quantity used to set the scale and the quantity
that is supposed to be calculated cancel as well as possible. 
For example, the lattice scale can be
determined using the static quark potential, which has small discretization errors ($O(a^2)$ or higher). 
A typical length scale is $r_0$ related to  the interquark force \cite{sommer1994} with
$r_0^2 \left.\frac{dV}{dr}\right|_{r=r_0} = 1.65$, which can be calculated on the lattice with 
high precision~\cite{guagnelli1998}.
The physical values corresponding to potential models are around $r_0 = 0.49-0.5$ fm. 
Unless noted otherwise we use $r_0=0.5$ fm.
For the string tension $\sigma$, usually experimental values of $\sqrt{\sigma} = 
427$ or 440 MeV are assumed. Other quantities frequently used to set the scale are the $\rho$ meson 
mass $m_\rho$, the nucleon mass, the decay constants $f_\pi$ and $f_K$, and charmonium and bottomonium level
splittings.

In Fig.~\ref{fig:aquenched} we show examples for the discrepancy between lattice spacings from
different physical quantities in the quenched approximation.
\begin{figure}[htb]
\vspace{0.1cm}
\begin{center}
\centerline{
\epsfysize=4.9cm \epsfbox{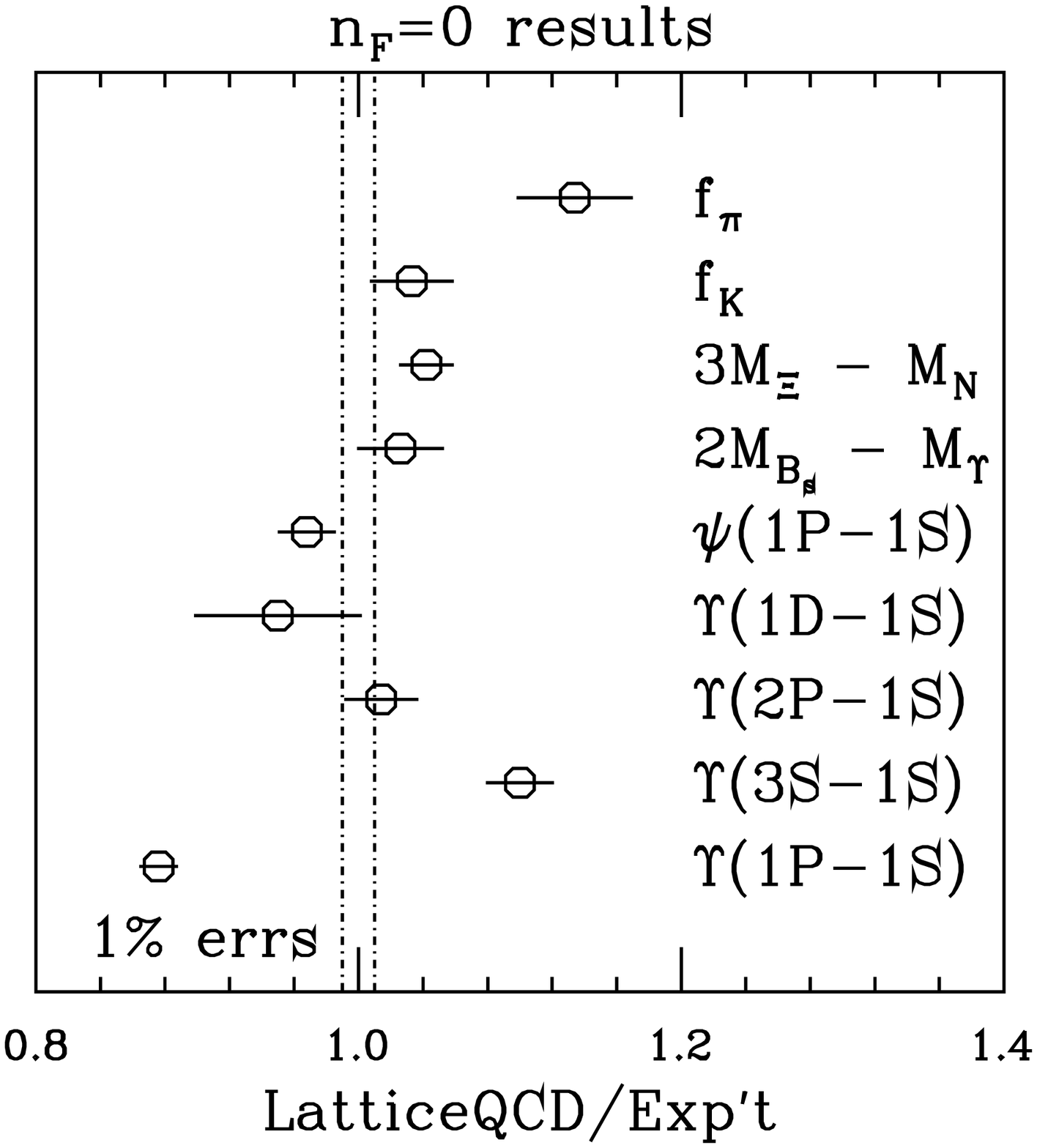}
\epsfysize=4.9cm \epsfbox{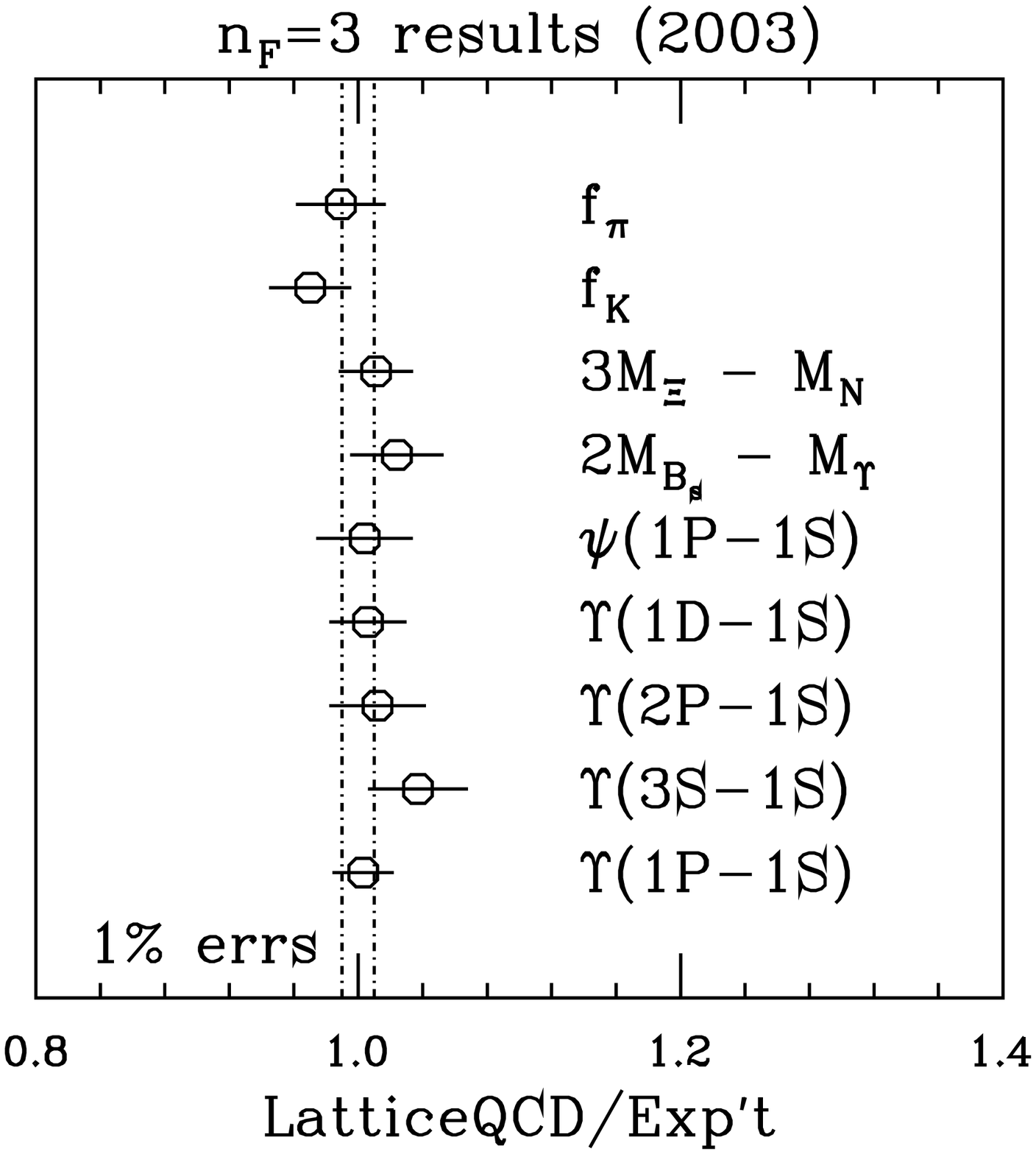}
}
\end{center}
\vspace{-0.5cm}
\caption{Comparison of lattice with experimental results from~\protect\cite{gottlieb2003}, 
using zero (left) and $N_f = 
2_{\mathrm{light}} +1_{\mathrm{strange}}$ flavors (right). $a$ from the $\Upsilon^\prime-\Upsilon$ mass splitting.}
\label{fig:lat_milc} 
\end{figure}

With two flavors ($N_f = 2$), the agreement is improved:
using Wilson gauge fields and two flavors of $O(a)$ improved clover sea quarks, Ref.~\cite{jlqcd2003} 
quotes an agreement of scales  from $m_\rho$,  $f_K$ and $r_0 = 0.5$ fm. 
However, in the two flavor calculations of~\cite{davies1997,collins1999} (Wilson gauge fields, 
staggered sea and clover valence quarks) and of~\cite{manke2000,cppacs2002hadr} 
(RG gauge fields and tadpole-improved clover sea and valence quarks) at $a \sim 0.5$ GeV$^{-1}$, 
a $\sim 20\%$ discrepancy between lattice spacings from $\chi_b-\Upsilon$ mass splittings
and $m_\rho$ remains.
\begin{figure}[htb]
\begin{center}
\centerline{
\epsfysize=4.3cm \epsfbox{cppacs_lh.eps}
\epsfysize=4.3cm \epsfbox{cppacs_lh_phi.eps}
\epsfysize=4.2cm \epsfbox{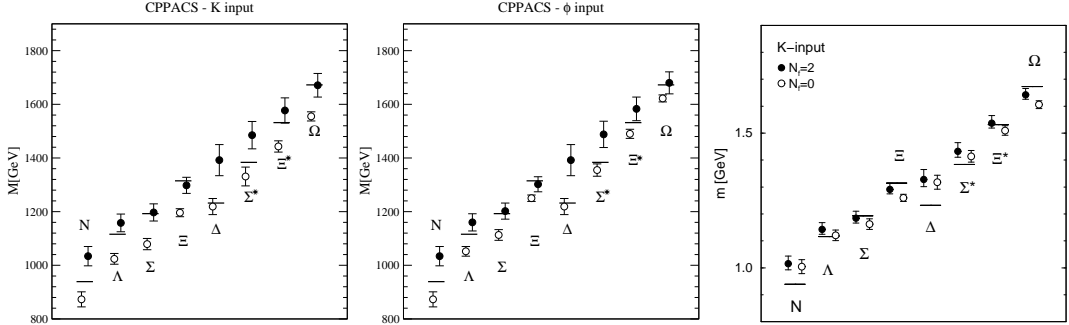}
}
\end{center}
\caption{Unquenched light baryon spectrum from the lattice.
Left and middle: results from ~\protect\cite{cppacs2002hadr}. Right figure:
from~\protect\cite{jlqcd2003hadr}. Open symbols denote
quenched, filled symbols unquenched results. $K$ input: strange quark mass set by fixing
the $K$ meson mass to the physical value, $\phi$ input: strange quark mass set by fixing the $\phi$ 
meson mass.}
\label{fig:bary_cppacs} 
\end{figure}

With two flavors of light and one flavor of strange dynamical quarks, using a 1-loop Symanzik $O(a^2)$ 
improved gauge action and a tree-level tadpole $O(a^2)$ improved staggered sea quark action 
Ref.~\cite{gottlieb2003} finds an agreement of a variety of physical quantities with experiment 
(see Fig.~\ref{fig:lat_milc}).
\subsection{The light baryon spectrum from the lattice}
In quenched calculations it was found that the features of the experimental light hadron
spectrum are described well by the lattice~\cite{butler1994,cppacs_q2003}.
It is of interest to study whether unquenching improves the agreement.
In Fig.~\ref{fig:bary_cppacs} we plot the baryon spectrum from 
the recent unquenched simulations of~\cite{cppacs2002hadr,jlqcd2003hadr}.
Discrepancies with experiment of $\sim 2\sigma$ remain. A reason may be uncertainty in the
chiral extrapolation. Ref.~\cite{jlqcd2003hadr} assigns additional systematic errors of up to
25 MeV from the  chiral extrapolation uncertainty and the determination of $r_0$.
\begin{table}[htb]
\begin{center}
\begin{tabular}{|l|l|l|l|l|l|}
\hline
                     & scale    & $\Lambda-N$[MeV]  & $\Delta-N$[MeV] & $\Sigma^\ast-\Sigma$[MeV]
& $\Xi^\ast - \Xi$[MeV] \\
\hline
\multicolumn{6}{|c|}{$N_f = 0$} \\
\hline
\protect\cite{cppacs2002hadr}  & $m_\rho$ & 
$151(34)(^{28}_{0})$ & 346(41) &  $252(33)(^{0}_{10})$ & $247(26)(^{0}_{7})$ \\
\protect\cite{jlqcd2003hadr}   & $m_\rho$ & 
 $116(33)(^{26}_0)$   & $314(37)$ & $252(30)(^0_{10})$ & $250(23)(^0_{11})$ \\
\hline
\multicolumn{6}{|c|}{$N_f = 2$} \\
\hline
\protect\cite{cppacs2002hadr}  & $m_\rho$ & $124(49)(^2_0)$ & 358(68) & 
$288(60)(^0_2)$ & $279(56)(^2_0)$ \\
\protect\cite{jlqcd2003hadr}   & $m_\rho$ & $128(26)(^{16}_0)$ & $313(31)$ &
$248(27)(^0_5)$ & $246(23)(^0_5$ \\
\hline
\multicolumn{6}{|c|}{$N_f =  2_{\mathrm{light}} +1_{\mathrm{strange}}$} \\
\hline
\protect\cite{aubin2004}  & $\chi_b-\Upsilon$ &  & 293(54) &  &  \\
\hline
\multicolumn{6}{|c|}{Model calculations} \\
\hline
\protect\cite{capstick1986}     &    & 155  & 270 &  180(15)  & 200(15)   \\
\hline
\multicolumn{6}{|c|}{Experiment} \\
\hline
       &   & 177 &  294 & 191 & 215 \\
\hline
\end{tabular}
\end{center}
\vspace{-0.2cm}
\caption{Light baryon mass splittings. The first error is statistical,
the second is the difference from fixing
the strange quark mass using the $K$ or $\phi$ meson where applicable.
The quantity used to fix the lattice scale is indicated.}
\label{tab:light_split}
\end{table}
In Table~\ref{tab:light_split} we give results for light baryon mass splittings corresponding to the
results shown in Fig.~\ref{fig:bary_cppacs}. For the splittings, the agreement with experiment 
is at the $1-2\sigma$ level. A recent calculation \cite{aubin2004} with 
$N_f = 2_{{\mathrm{light}}}+ 1_{{\mathrm{strange}}}$ finds a $\Delta-N$ splitting which agrees well 
with experiment.
\subsection{Nucleon mass: chiral extrapolation and finite size effects \label{sec:FS}}
\begin{figure}[htb]
\begin{center}
\centerline{
\epsfysize=4.5cm \epsfbox{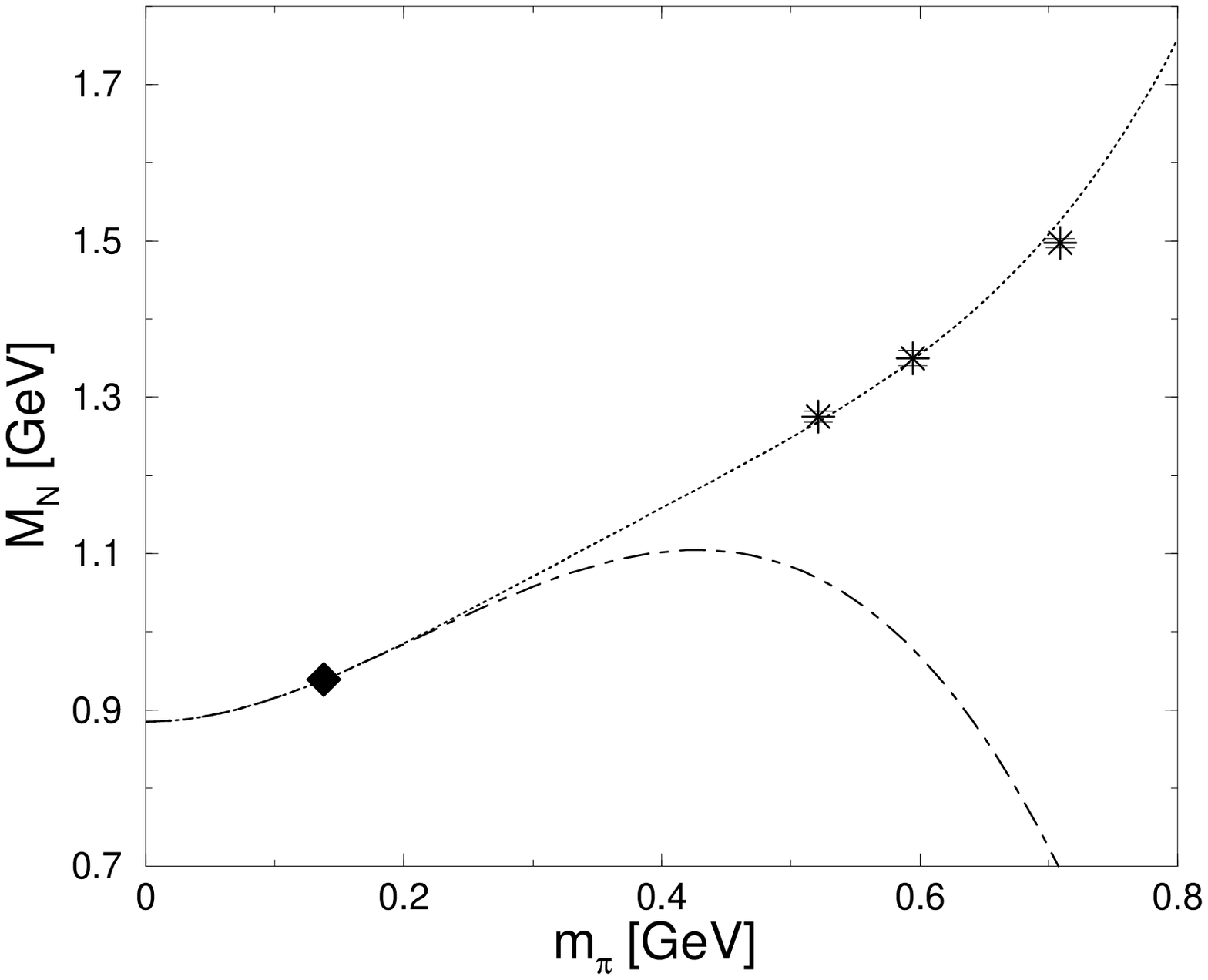}
\epsfysize=5.4cm \epsfbox{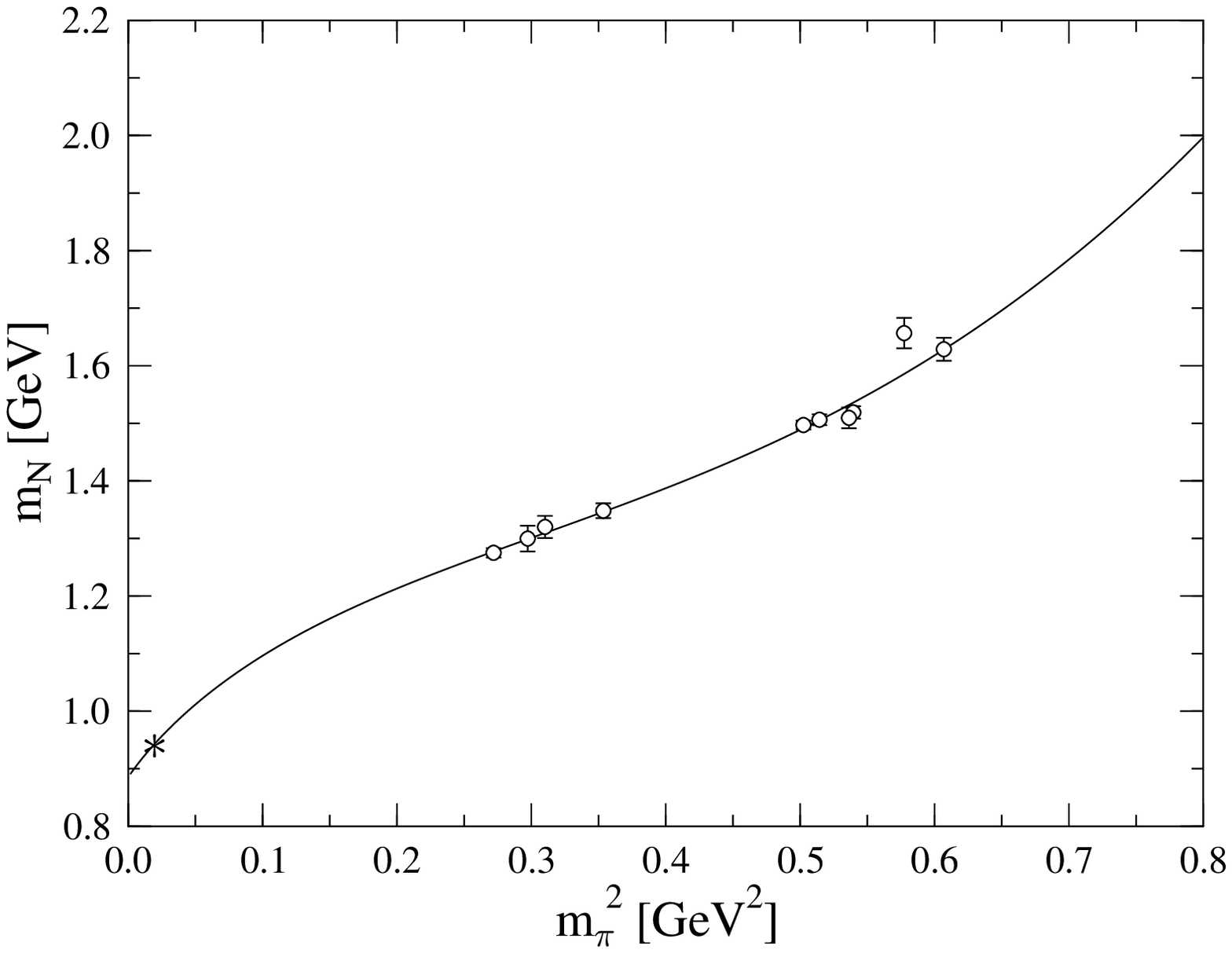}
}
\end{center}
\vspace{-0.3cm}
\caption{Chiral extrapolation of nucleon masses on large lattices
in two-flavor QCD using non-relativistic (left: from Ref.~\protect\cite{vbernard2004}) and relativistic 
(right: from Ref.~\protect\cite{meinulf2003}) $\chi PT$ at $O(p^4)$. The dot-dashed line on the left
shows the non-relativistic $O(p^3)$ result.}
\label{fig:mNfit} 
\end{figure}
$HB\chi PT$ predicts at $O(p^3)$ a correction   $\sim m_\pi^3$ to the quadratic dependence on 
$m_\pi$, but with a coefficient which is very different from the value found from fits to the 
lattice data. 
In Ref.~\cite{vbernard2004},  a good description of lattice data up to pion masses $\sim 600$ MeV could 
be achieved using the non-relativistic formalism at $O(p^4)$ (see Fig.~\ref{fig:mNfit} on the left).
With relativistic $\chi PT$  at $O(p^4)$~\cite{procura2004}, the agreement with 
the lattice data is also good up to rather large pion masses, as shown in  Fig.~\ref{fig:mNfit} on the 
right~\cite{meinulf2003}. 

\begin{figure}[htb]
\begin{center}
\centerline{
\epsfysize=6cm \epsfbox{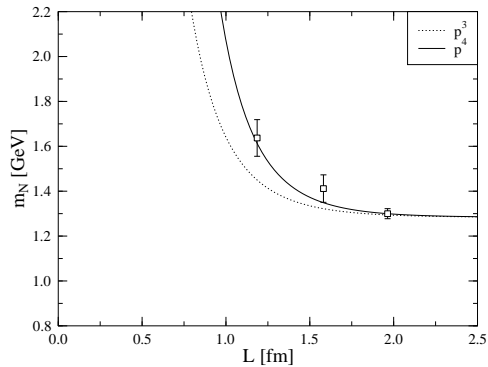}
}
\end{center}
\vspace{-0.7cm}
\caption{Volume dependence of the nucleon mass relativistic $\chi PT$ compared with
$N_f = 2$ lattice results, from Ref.~\protect\cite{meinulf2003}. Solid line: $O(p^4)$ result, 
dashed line: only $O(p^3)$ terms.}
\label{fig:FS}
\end{figure}

Having ensured that relativistic $\chi PT$ $O(p^4)$ indeed describes the nucleon mass on very large 
lattices, it should be possible to calculate the finite size effects on lattices which are not too small
within in this formalism.
Calculating the difference of the nucleon self-energy in a spatially finite and infinite volume 
within $\chi PT$ at $O(p^4)$~\cite{meinulf2003}, assuming an infinite temporal extent of the lattice, one finds 
a good agreement with the finite size behavior of the lattice results.
An example for a pion mass around 550 MeV is given in Fig.~\ref{fig:FS}.

The non-relativistic formalism at $O(p^3)$ predicts
finite size effects which are clearly smaller than the finite size effects of the lattice data~\cite{alikhan2002}.  
\subsection{The spectrum of hadrons with a $b$ quark}
A heavy quark with infinite mass can be regarded
as a color source which is static in the rest frame of the hadron and whose spin is not relevant to the
interactions.  Corrections due to the finiteness of the heavy quark mass can be included in a $1/M$ 
expansion.  Within the Heavy Quark Effective Theory (HQET), the mass of a heavy-light hadron $H$ can be
thought of as consisting of the following contributions:
\be
M_H = M_Q + \overline{\Lambda} - \frac{1}{2M_Q}\left[ \frac{\langle H|\overline{Q} \vec{D}^2 Q|H\rangle}{2M_H}
+ \frac{\langle H|\overline{Q} \vec{\sigma}\cdot\vec{B}Q|H\rangle }{2M_H}\right] + O(1/M_Q^2),
\label{eq:HQET}
\ee
where $Q$ is the heavy quark spinor, $M_Q$ the heavy quark mass, 
$\overline{\Lambda}$ the binding energy
of the meson for $M_Q \rightarrow \infty$, and the other two terms the
expectation values of the heavy quark kinetic energy and the spin-colormagnetic
interaction energy respectively.
We give a brief summary of lattice results on the hyperfine splittings $B^\ast-B$ and 
${B_s}\hspace{-0.1cm}^\ast-B_s$ in
Section~\ref{sec:hyp}. Results on $P$ wave states and baryons are given in Tables~\ref{tab:pstates},
\ref{tab:psplit} 
and~\ref{tab:bary}. Masses averaged over spin-orientations (spin-averaged) are denoted by an overbar.
The first error on the individual lattice results includes statistical errors and uncertainties
fixing the masses to the physical values, the second, where applicable, is a chiral extrapolation uncertainty.
To calculate weighted averages, we include an estimate of systematic errors from the actions. 
Most of the calculations use NRQCD, except for~\cite{bowler1996} who uses heavy clover quarks and
\cite{green2003} who simulates $B_s$ mesons in the static approximation
and interpolates between the static and experimental $D_s$ mesons.

Discretization 
errors with non-perturbatively $O(a)$ improved clover light quarks (finer lattice of~\cite{hein2000}
and~\cite{aoki2003}) are $O(a^2\lqcd^2)$, whereas the tadpole-improved light
clover action has $O(a^2\lqcd^2)$ and $O(\alpha_s a\lqcd)$ errors (coarser lattice of~\cite{hein2000}
and~\cite{alikhan2000,cppacs2000}). Refs.~\cite{lewis2000,mathur2002} use $O(a^2)$ 
tree-level tadpole-improved clover light actions respectively. Ref.~\cite{wing2003.stag} uses staggered
light quarks. 
The scale has been set with $m_\rho$ except in Ref.~\cite{cppacs2000} sets the scale with $\sqrt{\sigma} = 427$ 
MeV. Ref.~\cite{green2003} and ~\cite{burch2004} uses $r_0$ with physical values of 0.5 and 0.525 
fm respectively, and~\cite{wing2003.stag} uses $r_0 = 0.5$ fm and 
quarkonia at and around the charm~\cite{alford1998}. All other calculations from the set discussed here
use $m_\rho$.
The NRQCD action has errors $O(\alpha_s\lqcd/M)$ from  corrections to the spin-magnetic coefficient.
Errors on spin splittings are treated as being dominated by an error on the spin-magnetic coefficient of
$O(\alpha_s) \sim 20-30\%$.

The systematical error of each result is divided 
into a part common to all calculations, which is taken to be of the order of the error of the calculation 
with the smallest uncertainty, and and a rest which is treated as independent. 
The error on the average is rescaled by $r = \sqrt{\chi^2/(N-1)}$, where $N$ is the number of 
results, if $r > 1$. 
The second error on the averages comes from the variation due to the $10\%$ ambiguity between using $a$ from 
$m_\rho$ and $a$ from $r_0 = 0.5\mbox{ fm}$ in the quenched case, and asymmetric chiral 
extrapolations where applicable. 
The $\chi_b-\Upsilon$ mass difference is not included in the estimate of the scale variation
since it gives values for spin-independent mass splittings which are much higher than experiment. 
For example, Ref.~\cite{collins1999} quotes 
 $B_s-B$ and the $\Lambda_b-B$ splittings of 118 and 670 MeV if the scale is 
set with the $\chi_b-\Upsilon$ splitting instead with $m_\rho$. The experimental values are $90$ and $345$ MeV
respectively~\cite{pdg}.

If a collaboration quotes results from
several lattice spacings, they are plotted starting from the coarsest lattice on the left.
Asymmetric errors are added linearly in the plots.

For the error estimates we use  nominal values of $\lqcd = 400$ MeV,  $M = 5$ GeV and
$\alpha_s = \alpha_V(1/a)$, 
where $\alpha_V(q^\ast)$ is  defined in the potential scheme described in \cite{lepage1993} at the scale
$q^\ast$. Since 
the lattice results have rather varying central values we do not calculate the error in percent of the
individual lattice splittings but of the experimental splittings or nominal estimates thereof.
\subsubsection{$B^\ast-B$ splitting\label{sec:hyp}}
Results from quenched lattice NRQCD calculations of the $B^\ast-B$ and $B^\ast_s-B_s$ 
splittings fixing the scale with $m_\rho$ (e.g.\cite{hein2000,alikhan2000,cppacs2000,mathur2002,ishikawa2000}), 
and $r_0=0.5$ fm~\cite{wing2003.stag} are found to be around $25-35$ MeV, compared to experimental values of 
$45.8(4)$ MeV and $47.0(2.6)$ MeV~\cite{pdg}, respectively. 
Using relativistic $O(a)$ improved heavy quarks, Ref.~\cite{bowler2001} obtains a splitting 
around $10-20$ MeV with an error of 10 MeV.
A quenched calculation using the FNAL action~\cite{mackenzie1998} setting $a$ with the charmonium $1P-1S$ 
splitting quotes a $B^\ast_s-B_s$ splitting of around $40(10)$ MeV.
A preliminary calculation comparing results with zero and two flavors of $3\times$ the strange quark 
mass~\cite{cppacs2000} on lattices with $a \sim 0.2$ fm
finds an unquenched value of around 33 MeV with only an insignificant
increase compared to the quenched result from coarse lattices. A recent NRQCD calculation with $2+1$
dynamical flavors~\cite{wing2002.stag}
finds a $B^\ast_s-B_s$ splitting of $42.5(3.7)$ MeV.
\subsubsection{Orbitally excited $B$ mesons}
\begin{figure}[tbh]
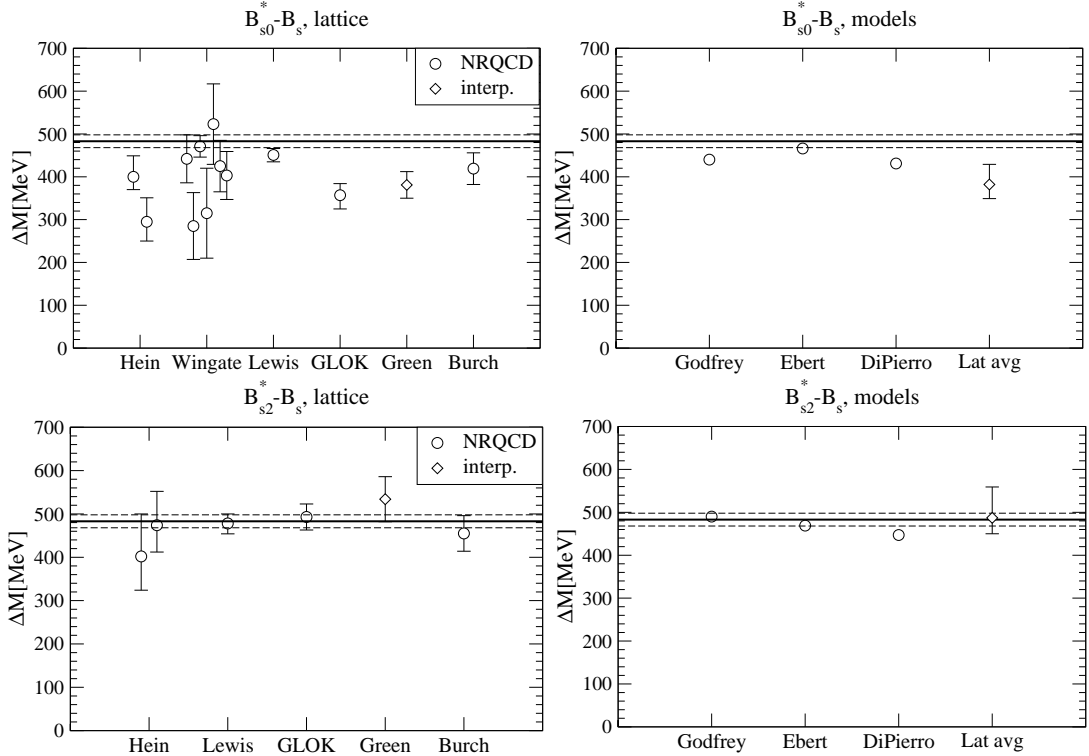

\begin{center}
\centerline{
\epsfysize=5cm \epsfbox{Bs0star-Bs-comp.eps}
\epsfysize=5cm \epsfbox{Bs0star-Bs-pot.eps}
}
\centerline{
\epsfysize=5cm \epsfbox{Bs2star-Bs-comp.eps}
\epsfysize=5cm \epsfbox{Bs2star-Bs-pot.eps}
}
\end{center}
\caption{Comparison of splittings between $P$ wave and the ground  state 
of $B_s$ mesons from Refs.~\protect\cite{hein2000,lewis2000,wing2003.stag,alikhan2000,green2003}
(lattice) and Refs.~\protect\cite{godfrey1985,ebert1998,dipierro2001} (models).
The lines denote the experimental value of the narrow $B_{sJ}^\ast(5850)$ resonance 
believed to come from orbitally excited $B_s$ mesons~\protect\cite{pdg}. }
\label{fig:Bsstarstar-Bs}
\end{figure}
\begin{table}[thb]
\begin{center}
\begin{tabular}{|l|ll|}
\hline
\multicolumn{1}{|c}{Ref.} &
\multicolumn{1}{c}{$\Delta M(B  )$[MeV]} &
\multicolumn{1}{c|}{$\Delta M(B_s)$[MeV]} \\
\hline
\hline
\multicolumn{3}{|c|}{$\bf{B_0^\ast-B}$} \\
\hline
\multicolumn{3}{|c|}{Lattice} \\
\hline
\protect\cite{hein2000}, $a \sim 1.1$GeV$^{-1}$  
                               &                         & $400(30)(^{19}_0)$ \\
\protect\cite{hein2000}, $a \sim 2.6$GeV$^{-1}$  
                                 &                         & $295(45)(^{11}_0)$ \\
\protect\cite{lewis2000}        &$475(^{19}_{20})$ &  $451(^{15}_{16})$ \\ 
\protect\cite{alikhan2000}      &  $374(37)$         &  $357(27)(^0_5)$  \\
\protect\cite{wing2003.stag}, 1l  &                   &  $442(56)$ \\
\protect\cite{wing2003.stag}, 1l  &                   &  $471(25)$ \\
\protect\cite{wing2003.stag}, 1l  &                   &  $315(105)$ \\
\protect\cite{wing2003.stag}, 1l  &                   &  $425(60)$ \\
\protect\cite{wing2003.stag}, Asq  &                   &  $285(78)$ \\
\protect\cite{wing2003.stag}, Asq  &                   &  $523(94)$ \\
\protect\cite{wing2003.stag}, Asq  &                   &  $403(56)$ \\
\protect\cite{green2003}        &                   &  $386(31)$ \\
\protect\cite{burch2004}        &  $408(67)$           &   $419(37)$ \\
average                         &  $402(41)(^{33}_7)$&  $382(23)(^{27}_{13})$      \\
\hline
\multicolumn{3}{|c|}{Model calculations} \\
\hline
\protect\cite{godfrey1985}      & 450                     & 440        \\
\protect\cite{ebert1998}        & 453                     & 466      \\
\protect\cite{dipierro2001}     & 427                     & 431   \\
\hline                        
\hline
\multicolumn{3}{|c|}{$\bf{B_2^\ast-B}$} \\
\hline
\hline
\multicolumn{3}{|c|}{Lattice} \\
\hline
\protect\cite{hein2000}, $a \sim 1.1$GeV$^{-1}$ 
                                &                         &$402(78)(^{20}_0)$ \\
\protect\cite{hein2000}, $a \sim 2.6$GeV$^{-1}$  
                               &                         & $474(62)(^{16}_0)$ \\
\protect\cite{lewis2000}       &   $493(^{29}_{32})$     & $478(^{22}_{24})$   \\
\protect\cite{alikhan2000}     &   $526(45)$             & $493(26)$   \\
\protect\cite{cppacs2000}      &   $426(17)$             &    \\
\protect\cite{green2003}       &                         & $534(52)$   \\
\protect\cite{burch2004}        &  $440(77)$             &  $455(41)$ \\
average                        &   $498(49)(^{42}_8)$    & $487(31)(^{44}_8)$ \\
\hline
\multicolumn{3}{|c|}{Model calculations} \\
\hline
\protect\cite{godfrey1985}     & 450                     & 490        \\
\protect\cite{ebert1998}       & 453                     & 469      \\
\protect\cite{dipierro2001}    & 435                     & 447  \\
\hline                        
\hline
\multicolumn{3}{|c|}{\bf{Preliminary experimental}} \\
\hline
\protect\cite{pdg} &     419(8) &  483(15)  \\
\hline
\end{tabular}
\end{center}
\caption{$B$ orbital excitations. Only statistical errors and, where quoted by the authors, 
errors due to chiral extrapolation and fixing the $b$ quark mass are shown. 
For the $B_2^\ast$ of \protect\cite{hein2000} at $a \sim 1.1$GeV$^{-1}$ we quote 
the result from the lattice operator they use for the $\overline{P}-\overline{S}$ splitting. 
The other lattice operator corresponding to the $B_2^\ast$ in their calculation
gives a $\sim 45$ MeV higher result.}
\label{tab:pstates}
\end{table}
\begin{table}[thb]
\begin{center}
\begin{tabular}{|l|ll|}
\hline
\multicolumn{1}{|c}{Ref.} &
\multicolumn{1}{c}{$\Delta M(B  )$[MeV]} &
\multicolumn{1}{c|}{$\Delta M(B_s)$[MeV]} \\
\hline
\hline
\multicolumn{3}{|c|}{$\bf{B_2^\ast-B_0^\ast}$} \\
\hline
\hline
\multicolumn{3}{|c|}{Lattice} \\
\hline
\protect\cite{hein2000}, $a \sim 1.1$GeV$^{-1}$  
                                &                         & $41(94)(^{14}_{0})$  \\
\protect\cite{hein2000}, $a \sim 2.6$GeV$^{-1}$  
                           &                         & $179(65)(^{6}_{0})$  \\
\protect\cite{lewis2000}   &   $18(^{36}_{38})$     & $27(^{27}_{29})$   \\
\protect\cite{alikhan2000} &   $155(32) $             & $136(23)$   \\
\protect\cite{green2003}   &                         & $148(61)$   \\
\protect\cite{burch2004}   &  $32(87)$               & $36(55)$     \\
average              &        $98(47)(^{10}_0)$            & $101(25)(^{11}_0)$ \\
\hline
\multicolumn{3}{|c|}{Model calculations} \\
\hline
\protect\cite{godfrey1985}      & 40                      & 50        \\
\protect\cite{isgur1998}        & $-155$                  &          \\
\protect\cite{ebert1998}        & $-5$                    & 3      \\
\protect\cite{dipierro2001}     & 7                       & 16  \\
\hline
\end{tabular}
\end{center}
\caption{$P$ state fine structure of $B$ mesons. Only statistical errors and errors due to chiral extrapolation 
and fixing the $b$ quark mass are shown.}
\label{tab:psplit}
\end{table}
\begin{table}[thb]
\begin{center}
\begin{tabular}{|l|l|l|}
\hline
\hline
\multicolumn{1}{|c}{Ref.} &
\multicolumn{1}{c}{$\Lambda_b-\overline{B}$[MeV]} &
\multicolumn{1}{c|}{$\overline{\Sigma}_b-\Lambda_b$[MeV]} \\
\hline
\hline
\multicolumn{3}{|c|}{{\bf Lattice}} \\
\hline
\protect\cite{bowler1996}                   &  $338(^{61}_{52})$    & $186(^{61}_{76})$  \\
\protect\cite{alikhan2000}                  &  $370(67)$              & 221(71)   \\ 
\protect\cite{mathur2002}, $a_s=1.1$GeV$^{-1}$
                                            &  $361(^{103}_{108})$  & $156(^{39}_{33})$  \\
\protect\cite{mathur2002}, $a_s=0.9$GeV$^{-1}$
                                            &  $367(^{108}_{110})$  & $191(^{38}_{37})$  \\
\protect\cite{aoki2003}                     &  $389(44)$            & 122(65)            \\
\protect\cite{cppacs2000}, quenched         &  $361(22)$                       &  \\
\protect\cite{collins1999}, $N_f=2$         &  $545(40)(22)$         &  \\
\protect\cite{cppacs2000}, $N_f=2$          &  $417(19)$                       &  \\
quenched average                            &  $372(33)(^{33}_5)$         &  $174(28)(^{15}_2)$  \\
\hline
\multicolumn{3}{|c|}{{\bf Models}} \\
\hline
\protect\cite{jenkins1992}                   & 312 & 196     \\ 
\protect\cite{capstick1986}            &     & 217     \\
\protect\cite{capstick1986}$c$ quark   &     &  212     \\
\hline
\multicolumn{3}{|c|}{{\bf Experiment}} \\
\hline
\protect\cite{pdg}                       & 311(10) &               \\
\hline
\hline
\multicolumn{1}{|c}{Ref.} &
\multicolumn{1}{c}{$\Sigma_b^\ast-\Sigma_b$[MeV]} &
\multicolumn{1}{c|}{$\Omega_b^\ast-\Omega_b$[MeV]} \\
\hline
\hline
\multicolumn{3}{|c|}{{\bf Lattice}} \\
\hline
\protect\cite{alikhan2000}        &    $19(7) $           & $18(4)$   \\
\protect\cite{mathur2002}, $a_s=1.1$GeV$^{-1}$  
                                   & $22(12)$           &  $18(^9_8)$ \\
\protect\cite{mathur2002}, $a_s=0.9$GeV$^{-1}$  
                                   & $24(^{13}_{12})$           & $20(9)$  \\
\protect\cite{aoki2003}            & 10(12)            & 7(4)    \\
average                            & $18(8)(^4_0)$     & $13(7)(^3_0)$  \\
\hline
\multicolumn{3}{|c|}{{\bf Models}} \\
\hline
\protect\cite{capstick1986}           &  10  &                 \\
\protect\cite{jenkins1992}   & 8           &     \\
\hline
\end{tabular}
\end{center}
\caption{$b$ baryons. Only statistical errors and, where applicable, systematical errors due to scale 
setting except for the quenched scale ambiguity and fitting are shown.}
\label{tab:bary}
\end{table}
For heavy-light mesons it seems appropriate to use a hydrogen-like picture for the coupling of 
angular momenta of the quarks. In the infinite mass limit, there are 
two $P$ wave energy levels with 
light quark angular momentum $j_l = 1/2$ and $3/2$. At finite $M$ there is an additional hyperfine
structure due to the coupling of the heavy quark spin. This results in one level  with 
angular momentum zero ($B_0^\ast$), two with angular momentum one ($B_1^\prime$ and $B_1$), and
one with angular momentum two ($B_2^\ast$). Here we discuss the $B_0^\ast$ and $B_2^\ast$ level
splittings. 

Lattice results are presented in Table~\ref{tab:pstates}.
In the error estimation,  we use guessed values where no experimental value 
is available: 400 MeV for $B_0^\ast-B$ and $B_{s0}^\ast-B_s$ and
500 MeV for $B_2^\ast-B$ and $B_{s2}^\ast-B_s$.

Experimental knowledge of the $P$ state level structure is still sparse. The Particle Data
Book lists two candidates. 
There is the $B_J^\ast(5732)$ resonance with a width 
of $\sim 130$ MeV, which is believed to come from several narrow and broad 
$P$ wave states, and the 
$B_{sJ}^\ast(5850)$ signal with a width of $\sim 50$ MeV, which can be interpreted as stemming from
excited $B_s$ states. A comparison of the lattice results with experiment and
with model calculations is given in Fig.~\ref{fig:Bsstarstar-Bs}.

The sign of the $B_2^\ast-B_0^\ast$ mass difference is disputed among 
potential model calculations (e.g. \cite{godfrey1985,isgur1998,ebert1998,dipierro2001}).
Individual lattice calculations~\cite{hein2000,lewis2000} find a splitting around zero and
are within errors compatible with a small negative splitting, but the lattice average for 
$B_2^\ast-B_0^\ast$ is positive.
\begin{figure}[thb]
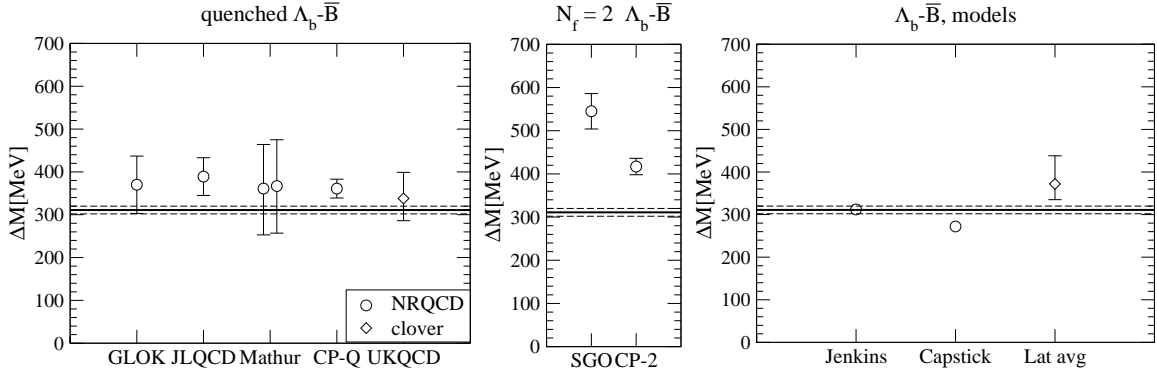

\begin{center}
\centerline{
\epsfysize=4.9cm \epsfbox{lambdab-comp.eps}
\epsfysize=4.9cm \epsfbox{lambdab-comp-dyn.eps}
\epsfysize=4.9cm \epsfbox{lambdab-pot.eps}
}
\end{center}
\vspace{-0.5cm}
\caption{$\Lambda_b-\overline{B}$ splitting from  quenched (left) and
$N_f = 2$ (middle) lattices 
(Refs.~\protect\cite{collins1999,alikhan2000,cppacs2000,mathur2002,aoki2003,bowler1996}). 
The lattice average shown on the right is quenched. CP-Q and CP-2
denote quenched and unquenched results from \protect\cite{cppacs2000}, respectively. Model results are from 
Refs.~\protect\cite{capstick1986,jenkins1992}.}
\label{fig:lambdab-comp}
\end{figure}
\subsubsection{$b$ baryons}
Baryons with one $b$ quark can be thought of as two light quarks coupling to form a spin zero or 
spin one diquark. The state with a spin zero diquark is the $\Lambda_b$. If the diquark has spin one, 
the heavy quark can couple to a spin $1/2$ state, the $\Sigma_b$, and a spin $3/2$ state, 
the ${\Sigma_b}\hspace{-0.1cm}^\ast$. If the light quarks in the $\Sigma_b$ and the 
${\Sigma_b}\hspace{-0.1cm}^\ast$ 
are substituted by strange quarks one obtains the $\Omega_b$ and the ${\Omega_b}\hspace{-0.1cm}^{\ast}$.

In Table~\ref{tab:bary} we summarize results for 
the spin-independent splittings $\Lambda_b-\overline{B}$ and 
$\overline{\Sigma}_b-\Lambda_b$ with $M(\overline{\Sigma}_b) = [2M(\Sigma_b) + 
4M({\Sigma_b}\hspace{-0.1cm}^\ast)]/6$ and $M(\overline{B}) = [3M(B^\ast)+M(B)]/4$, and the 
${\Sigma_b}\hspace{-0.1cm}^\ast-\Sigma_b$ 
hyperfine splitting. The values quoted for \cite{aoki2003} are obtained by interpolating the 
heavy-strange meson mass to the $B_s$ mass and interpolating the baryon splittings to the thus 
obtained $b$ quark mass.

We compare lattice results with calculations within a constituent quark model~\cite{capstick1986}
and a Skyrme model~\cite{jenkins1992} where the
baryon is described as a bound state between a soliton and a heavy quark.

The experimental values of the $\Lambda_b-\overline{B}$  and the $\Lambda_c-\overline{D}$ 
splittings are very close: 311(9) and 310(2) MeV respectively.  The only $1/M$ correction to these
splittings comes from the heavy quark kinetic energy. Its contribution to the $\Lambda_b$ mass
appears to be very close to the contribution to the $B$ mass.

The quenched lattice average differs from the experimental value by less than $2 \sigma$.
Preliminary results with two flavors of dynamical quarks around the 
strange quark mass~\cite{collins1999,cppacs2000} are even higher. Ref.~\cite{cppacs2000}
finds an increase of $\sim 15\%$ if $N_f$ is changed from zero to two dynamical
quarks of around  $3\times$ the strange quark mass.
\begin{figure}[thb]
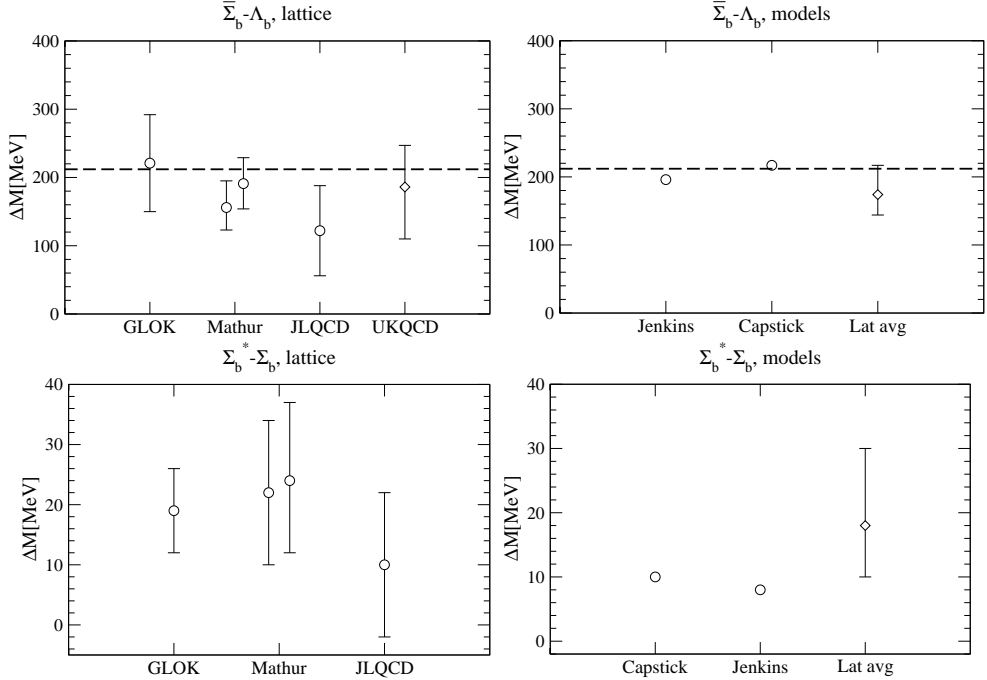

\begin{center}
\centerline{
\epsfysize=4.5cm \epsfbox{sigmab-lambdab-comp.eps}
\epsfysize=4.5cm \epsfbox{sigmab-lambdab-pot.eps}
}
\centerline{
\epsfysize=4.5cm \epsfbox{sigmastar-sigma-comp.eps}
\epsfysize=4.5cm \epsfbox{sigmastar-sigma-pot.eps}
}
\end{center}
\vspace{-0.5cm}
\caption{Above: $\overline{\Sigma}_b-\Lambda_b$ splitting from the lattice (left) and model calculations
(right). The dashed line shows the
experimental  value for the $\overline{\Sigma}_c-\Lambda_c$ splitting. Below: 
${\Sigma_b}\hspace{-0.1cm}^\ast-\Sigma_b$ splitting. Lattice data from 
Refs.~\protect\cite{alikhan2000,mathur2002,aoki2003,bowler1996}, model
results from Refs.~\protect\cite{capstick1986,jenkins1992}.}
\label{fig:sigmab-comp}
\end{figure}
Calculation at smaller quark masses would clarify whether the difference is related to a
chiral extrapolation uncertainty.

In Fig.~\ref{fig:sigmab-comp} we show results for the spin-averaged $\overline{\Sigma}_b-\Lambda_b$ 
splitting.
Ref.\cite{jenkins1992} gives a relation between the heavy-light and light baryon splittings,
$\Delta M\left(\overline{\Sigma}_Q - \Lambda_Q\right)/\Delta M\left(\Delta-N\right) = 2/3$,
for $Q = c,b$. For $Q=c$, the equality holds well experimentally. 
Lattice results for the ratio with $Q=b$ vary between 0.5 and 1. 

In Fig.~\ref{fig:sigmab-comp} we also give results for the ${\Sigma_b}\hspace{-0.1cm}^\ast
-\Sigma_b$ hyperfine splitting. The expectation from HQET is that the hyperfine splitting is generated 
by the spin-chromomagnetic interaction (see Eq.~(\ref{eq:HQET})) and should  be
proportional to $1/M_Q$. Rescaling the experimental value of $\Delta M({\Sigma_c}\hspace{-0.1cm}^\ast-
\Sigma_c) \simeq 67$ MeV by the ratio of $b$ and $c$ quark masses, one expects 
$\Delta M({\Sigma_b}\hspace{-0.1cm}^\ast-\Sigma_b) \sim 20$ MeV, and the lattice results are compatible with
this expectation.  The error on the 
${\Sigma_b}\hspace{-0.1cm}^\ast-\Sigma_b$ and ${\Omega_b}\hspace{-0.1cm}^\ast-\Omega_b$ splittings is estimated
to be $O(\alpha_s) \times 20$ MeV.
\subsection{$f_B$}
In Table~\ref{tab:fB} we summarize lattice results for $f_B$ and $f_{B_s}$ since 1998. The first error given in the 
Table is statistical, and the second is the systematical error given by the authors added in quadrature.

First we address unquenching effects on $f_B$. They depend on how the scale is set. 
In Table~\ref{tab:rat} we compare ratios of decay constants from quenched and two-flavor simulations with the same 
gauge field and valence quark actions and find an increase of $\sim 10\%$ if $a$ is set with $f_\pi$,
$10-20\%$ if $a$ is set with $m_\rho$ and no increase with $\Upsilon$ and (for $f_{D_s}$) with $r_0$.

We calculate weighted averages for  quenched,  $N_f = 2$ and $N_f = 2+1$ results of $f_B$ and
$f_{B_s}$. Since the methods for error estimation can vary considerably between different collaborations, 
even if  similar lattice actions and parameters are used,
we make new assignments motivated by the error analysis of the authors themselves.
We assign common systematic errors to the
calculations with RG gauge fields using NRQCD~\cite{cppacs2001NR} and using the FNAL heavy quark action
as non-relativistic effective field theory without taking the continuum 
limit~\cite{cppacs2001R}, and to the calculations using Wilson gauge fields and
NRQCD (quenched are~\cite{alikhan1998,ishikawa2000,collins2001}). 
For quenched configurations with clover light quarks, 
we assign systematic errors of 20 and 22 MeV respectively for $f_B$  and $f_{B_s}$.
Ref.~\cite{wing2003.stag} uses NRQCD with an $O(a^2)$ tadpole improved gauge action and staggered light 
valence quarks, and we use their own systematic error assignment.

The quenched calculations of Refs.~\cite{bowler2001,abada2000,lin2001,becirevic2001} use Wilson gauge and
clover quarks at $a = 0.35-0.37$ GeV$^{-1}$ ($\beta = 6.2$)  simulated at the charm quark mass. 
Ref.~\cite{lin2001} uses tree-level tadpole-improved clover quarks without including $O(\alpha_s \times a)$
terms in the renormalization.
Ref.~\cite{abada2000} uses a non-perturbatively $O(a)$ improved clover quark action and a partly  non-perturbative
current renormalization.
Refs. \cite{bowler2001,becirevic2001} use nonperturbative $O(a)$ improvement except for a
perturbative value for the $O(\alpha_s am_q)$ quark mass correction to the renormalization constant.
Although different degrees of improvement are used, and the scaling behaviour is found to be different,
the results for $f_{D_s}$ agree at $\beta = 6.2$. We therefore assign a common systematic error to these results.
According to the estimate of the discretization
error given in~\cite{lin2001}  (8\%) and of a $1/M$ extrapolation error of $\sim 9\%$
given in~\cite{bowler2001} we use 23  MeV for $f_B$ and 26 MeV for $f_{B_s}$.

Refs.~\cite{aoki1998,elkhadra1998,bernard2002} use heavy quarks in the FNAL formalism and extrapolate
their results to $a \rightarrow 0$.
Refs.~\cite{guagnelli2002,dedivitiis2003} use a step scaling method with the Schr\"odinger functional 
and clover heavy quarks. Part 
of the renormalization factors is calculated nonperturbatively. 
 Their results are continuum extrapolated. Ref.~\cite{dellamorte2004} uses an interpolation 
between static and clover charm quarks which are non-perturbatively improved using the Schr\"odinger
functional and continuum extrapolated.

\begin{table}[thb]
\begin{center}
\begin{tabular}{|l|c|ll|}
\hline
\hline
\multicolumn{1}{|l}{Ref.} &
\multicolumn{1}{c}{scale} &
\multicolumn{1}{c}{$f_B$[MeV]} & 
\multicolumn{1}{c|}{$f_{B_s}$[MeV]} \\
\hline
\hline
\multicolumn{4}{|c|}{{\bf Lattice}} \\
\hline
\multicolumn{4}{|c|}{$N_f = 0$} \\
\hline
\protect\cite{aoki1998}  & $m_\rho$     &  $173(4)(13)$ & $199(3)(14)$ \\
\protect\cite{elkhadra1998}  & $f_K$   &  $164(^{14}_{11})(8)$ & $185(^{13}_{8})(9)$ \\
\protect\cite{alikhan1998}   & $m_\rho$ &  $147(11)(^{13}_{16})$          & $175(8)(16)$ \\
\protect\cite{ishikawa2000} & $\sqrt{\sigma}=427$MeV & 170(5)(15) & 191(4)(17) \\
\protect\cite{abada2000} &$\frac{M_K^\ast}{M_K},M_K^\ast$  & $173(13)(^{34}_2)$ & $196(11)(^{42}_0)$\\
\protect\cite{bowler2001}   & $f_\pi$   & $195(6)(^{23}_{24})$ & $220(6)(^{23}_{28})$ \\
\protect\cite{lin2001}       & $f_K$    & $177(17)(22)$  & $204(12)(^{24}_{23})$ \\
\protect\cite{collins2001}  & $m_\rho$  &                      & 187(4)(15)    \\
\protect\cite{cppacs2001R} & $m_\rho$   & 188(3)(26)           & $220(2)(^{32}_{31})$ \\
\protect\cite{cppacs2001NR} & $m_\rho$  & 191(4)(27)           & 220(4)(31) \\
\protect\cite{becirevic2001} &$\frac{M_K^\ast}{M_K},M_K^\ast$ & $174(22)(^{8}_{0})$& $204(15)(^8_0)$\\
\protect\cite{bernard2002}  & $f_\pi$   & $173(6)(16)$ & $199(5)(^{23}_{22})$ \\
\protect\cite{guagnelli2002}    & $r_0$    & 170(11)(23)      & 192(9)(25)   \\
\protect\cite{dedivitiis2003}   & $r_0$    &                  & 192(6)(4) \\
\protect\cite{dellamorte2004} & $r_0$    &                  & 205(12) \\
\protect\cite{wing2003.stag} & $r_0$    &                     & 225(9)(34)      \\
        average1       &  &  $175(7)(^{48[21]}_{4})$        &     $198(5)(^{46[9]}_{16})$            \\
        average2      &  &                                 &     $201(6)(^{51[13]}_{13})$            \\
\hline
\multicolumn{4}{|c|}{$N_f = 2$} \\
\hline
\protect\cite{collins1999}  & $m_\rho$   & 186(5)(25)               & $215(3)(^{28}_{29})$ \\
\protect\cite{cppacs2001R} & $m_\rho$    & 208(10)(29)              & $250(10)(^{36}_{35})$ \\
\protect\cite{cppacs2001NR} & $m_\rho$   & 204(8)(29)               & 242(9)(34) \\
\protect\cite{bernard2002}  & $f_\pi$    & $190(7)(^{25}_{17})$    & $217(6)(^{36}_{28})$ \\
\protect\cite{jlqcd2003}    & $m_\rho$   & $191(10)(^{10}_{22})$   & $215(9)(^{14}_{13})$ \\
\protect\cite{onogi2004}    & $r_0=0.49$fm &   $181(7)(^{20}_{29})$   &  \\
average1               &            &    $190(10)(^{55[6]}_{13})$        & $226(15)(^{53[7]}_{2})$          \\
average2               &            &                                     & $226(15)(^{55[8]}_{1})$              \\
\hline
\multicolumn{4}{|c|}{$N_f = 2+1$} \\
\hline
\protect\cite{wing2003}    & $\Upsilon (2S-1S)$        &       & 260(7)(28) \\
\hline
\hline
\multicolumn{4}{|c|}{{\bf Sum rules}} \\
\hline
\cite{narison2001} & & 203(23)  & 236(30) \\
\cite{penin2002}   & & 206(20)   &        \\
\cite{jamin2002} & & 210(19)  & 244(21) \\
\cite{braun1999}   & & $180-190(30)$  &         \\
\hline
\hline
\multicolumn{4}{|c|}{{\bf Potential models}} \\
\hline
\cite{ebert2002} & & 178(15)   & 196(20)  \\
\hline
\end{tabular}
\end{center}
\caption{$f_B$ and $f_{B_s}$ from the lattice. Statistical errors and systematical errors given by the authors 
are included, adding the systematical errors in quadrature. The method to set the scale is indicated in the 
second  column. The first error on the averages is due to the statistical and systematical errors of the individual 
results,
while the second error is from chiral extrapolation uncertainties and scale ambiguities as 
explained in the text. }
\label{tab:fB}
\end{table}

\begin{table}[htb]
\begin{center}
\begin{tabular}{|lcl||cl|}
\hline
\multicolumn{5}{|c|}{Ratios of decay constants} \\
\multicolumn{1}{|c}{Ref.} &
\multicolumn{1}{c}{scale} &
\multicolumn{1}{c|}{$f_B^{N_f=2}/f_B^{N_f=0}$} &
\multicolumn{1}{|c}{Ref.} &
\multicolumn{1}{c|}{$f_{B_s}/f_{D_s}$} \\
\hline
\cite{bernard2002} & $f_\pi$  & 1.10(6) & \cite{cppacs2001R}($N_f=0$) &  0.88(1) \\                      
\cite{bernard2002} & $m_\rho$ & 1.19(6) & \cite{bernard2002}($N_f=0$)  & $0.891(12)(^{40}_{34})$ \\  
\cite{cppacs2001R} & $m_\rho$ & 1.11(6) & \cite{dellamorte2004} ($N_f=0$)  &     0.81(6)       \\
\cite{cppacs2001NR}& $m_\rho$ & 1.07(5) &  \cite{cppacs2001R}($N_f=2$)  & 0.94(6)  \\                 
\cite{cppacs2001NR}& $\Upsilon(\overline{P}-\overline{S})$ & 
0.97(5)  &  \cite{bernard2002}($N_f=2$) &  $0.922(13)(^{68}_{55})$ \\
\cite{maynard2002}($f_{D_s}$) & $r_0$   & 0.98(4) &   & \\
\hline
\end{tabular}
\end{center}
\caption{Ratios of decay constants. The first error is statistical, the second the systematical
errors given by the authors added in quadrature, where applicable.}
\label{tab:rat}
\end{table}
The second error on the quenched results includes the  ambiguity between scales from 
$m_\rho$ and $\Upsilon$ level splittings by
varying the result by $+30\%$ if the scale is taken from $m_\rho$ or $\sqrt{\sigma}=427$ MeV, 
$(-3+27)\%$ if the scale is set with
$f_\pi$,  and $(-12+18)\%$ if the scale is set with $r_0 = 0.5$ fm. 
The lattice spacings calculated in \cite{lin2001,becirevic2001} from $K$ physics are
close to the results using $f_\pi$ at the same $\beta$ values, and the scales from $f_K$ 
determined in
\cite{elkhadra1998} are close to the ones using $m_\rho$ from the same actions and  $\beta$ values.
We also quote the upper bound from the variation between scales from $m_\rho$ and $r_0 = 0.5$ fm
(second error given in square brackets) which is slightly lower than but 
within errors compatible with the unquenched central value with $N_f = 2+1$. The uncertainty in fixing the 
strange quark mass is also included. Where it is not given by the authors, we include a $(+7)$
MeV error.

For two-flavor QCD, we estimate the systematic errors from non-relativistic 
methods~\cite{collins1999,cppacs2001R,cppacs2001NR,jlqcd2003} to be
21 MeV for $f_B$ and $25$ MeV for $f_{B_s}$. For the systematic error of the continuum extrapolated 
results with FNAL heavy quarks~\cite{bernard2002,onogi2004} we use the estimate of 
Ref.~\cite{bernard2002} including errors due to continuum extrapolation, perturbation theory, $1/M$ extrapolation
and, where applicable, the spin-magnetic coefficient.
 Ref.~\cite{jlqcd2003} makes an estimate of the light quark mass dependence of 
$f_B$ using 1-loop $\chi PT$ and quotes an uncertainty  $(^{0}_{19})$ MeV on $f_B$ from the
chiral extrapolation.
We assign the same error also to the $N_f = 2$ results of 
Refs.~\cite{collins1999,cppacs2001NR,cppacs2001R,onogi2004}.
For the values of Ref.~\cite{bernard2002} we use their own estimate of the chiral extrapolation error.
The uncertainty in the chiral extrapolations, a $\sim 50$ MeV increase in the decay constants if the 
$\chi_b-\Upsilon$ splitting  instead of $m_\rho$ is used to set the scale quoted by \cite{collins1999} and 
\cite{cppacs2001NR}, and the variation from determining the strange quark mass by setting the $K$ or $\phi$ 
meson mass to the physical value give the second error on the $N_f = 2$ averages.
The variation between using $m_\rho$ and $f_\pi$ to set the scale and  the chiral 
extrapolation uncertainty determined by \cite{bernard2002}, and the variation in the strange quark mass 
give the second error in square brackets.

The results of this procedure are included in Table~\ref{tab:fB} as average1.

Refs.~\cite{cppacs2001R} and~\cite{collins1999} quote  statistical errors on their quenched and 
$N_f = 2$ results on $f_{B_s}$ respectively which are about half the statistical errors of other calculations
with a similar ensemble size. 
We therefore also calculate the average with their statistical errors and the error of Ref.~\cite{dedivitiis2003} 
enlarged by a factor of two. The result, which is very close to average1, is quoted as average2 in 
Table~\ref{tab:fB}. Except for the uncertainties due to fixing the scale and reaching the physical quark masses, 
the errors on these averages are only few percent. 
The double ratio of $B$ and $D$ decay constants
$ f_{B_s}\sqrt{M_{B_s}}/(f_B\sqrt{M_B}\times f_D\sqrt{M_D}/(f_{D_s}\sqrt{M_{D_s}})$ 
should be independent of the exact form of the chiral extrapolation up to $1/M_Q$ corrections, as is supported by 
the results of a two-flavor calculation using FNAL heavy quarks~\cite{onogi2004}.
As argued by Ref.~\cite{becirevic2003.1} using $\chi PT$, the chiral extrapolation uncertainty of the ratio 
$f_{B_s}\sqrt{M_{B_s}}/(f_B\sqrt{M_B})\times f_\pi/f_K$ should also be small.

Employing unquenched staggered ($N_f=2+1$) MILC gauge field 
ensembles at $a \simeq 0.13$ fm, the NRQCD estimate of \cite{wing2003} is:
\be
f_{B_s} = 260(7)(28), \label{eq:res}
\ee
where systematical errors are added in quadrature.
Calculations of the decay constants with the staggered $N_f=2+1$ configurations using NRQCD~\cite{gray2004} and 
FNAL \cite{bernard2004} heavy quarks are in further progress. 
Within the statistical and systematical errors quoted in Table~\ref{tab:fB}, the results with 
$N_f = 0, 2$ and $2+1$ agree among each other.


We relate this to the experimental value for $f_{D_s}$ using unquenched lattice results for the ratio
$f_{B_s}/f_{D_s}$ from two-flavor calculations which work directly at the $b$ and $c$ quark masses 
without using extrapolations.

Taking the experimental value $f_{D_s} = 283(45)$ MeV (Eq.~(\ref{eq:fDs})), and  the range of
values for the ratio $f_{B_s}/f_{D_s}$ from Table~\ref{tab:rat}, one obtains 
$f_{B_s} = 230-260$ MeV.

Other recent review articles~\cite{ryan2002,lellouch2002,becirevic2003,wittig2003,hashimoto2004}
quote lattice estimates for $f_B$ and $f_{B_s}$ which are within errors in agreement with the
averages quoted in Table~\ref{tab:fB}.

In Table~\ref{tab:fB} we compare the lattice results with recent sum 
rule~\cite{narison2001,penin2002,jamin2002,braun1999} and potential model calculations~\cite{ebert2002},
and we find that they are within errors in agreement.
\section{Conclusions}
Applications of non-relativistic QCD and chiral perturbation theory in lattice calculations are presented.
The status of lattice results on the light and heavy-light hadron spectrum and the decay constants $f_B$ and 
$f_{B_s}$ is summarized, and weighted lattice averages for $b$ hadron mass splittings and 
decay constants are calculated. 
The agreement of the  hadron spectrum with experiment is a major success of lattice QCD in general, and
of non-relativistic methods for heavy quarks in particular, and supports 
the reliability of lattice predictions of hadronic matrix elements.
The lattice has become instrumental in  QCD calculations. 
Work on further understanding and reduction of lattice errors 
is in progress and will enable very precise checks.
\subsection*{Acknowledgements}
I thank D.~Ebert, R.N.~Faustov, V.O.~Galkin, A.~Sch\"afer and G.~Schierholz
for discussions and comments on the manuscript, and C.~Bernard, V.M.~Braun,  J.~Koponen, 
L.~Lellouch and C.~Michael for discussions. I thank D.~Toussaint for the numbers
for the unquenched light baryons from MILC (\cite{aubin2004}). 

I am grateful for a personal fellowship of the Deutsche Forschungsgemeinschaft. I would like to
thank the group ``Theory of Elementary Particles/Phenomenology-Lattice Gauge Theories'' of the 
Humboldt University Berlin and the NIC/DESY Zeuthen for their kind hospitality. 

The numerical computations relevant to my publications were performed at the computer centers 
EPCC (Edinburgh), LANL (Los Alamos), LRZ (M\"unchen), NIC (J\"ulich) and NIC (Zeuthen),
NCSA (Urbana-Champaign), RCCP (Tsukuba) and SCRI (Tallahassee).
I thank all institutions for their support.

\end{document}

%% file: epsf2.tex
\ifx\epsfannounce\undefined \def\epsfannounce{\immediate\write16}\fi
 \epsfannounce{This is `epsf.tex' v2.7k <10 July 1997>}%
\newread\epsffilein    
\newif\ifepsfatend     
\newif\ifepsfbbfound   
\newif\ifepsfdraft     
\newif\ifepsffileok    
\newif\ifepsfframe     
\newif\ifepsfshow      
\epsfshowtrue          
\newif\ifepsfshowfilename 
\newif\ifepsfverbose   
\newdimen\epsfframemargin 
\newdimen\epsfframethickness 
\newdimen\epsfrsize    
\newdimen\epsftmp      
\newdimen\epsftsize    
\newdimen\epsfxsize    
\newdimen\epsfysize    
\newdimen\pspoints     
\def\epsfbox#1{\global\def\epsfllx{72}\global\def\epsflly{72}%
   \global\def\epsfurx{540}\global\def\epsfury{720}%
   \def\lbracket{[}\def\testit{#1}\ifx\testit\lbracket
   \let\next=\epsfgetlitbb\else\let\next=\epsfnormal\fi\next{#1}}%
\def\epsfgetlitbb#1#2 #3 #4 #5]#6{%
   \epsfgrab #2 #3 #4 #5 .\\%
   \epsfsetsize
   \epsfstatus{#6}%
   \epsfsetgraph{#6}%
}%
\def\epsfnormal#1{%
    \epsfgetbb{#1}%
    \epsfsetgraph{#1}%
}%
\newhelp\epsfnoopenhelp{The PostScript image file must be findable by
TeX, i.e., somewhere in the TEXINPUTS (or equivalent) path.}%
\def\epsfgetbb#1{%
    \openin\epsffilein=#1
    \ifeof\epsffilein
        \errhelp = \epsfnoopenhelp
        \errmessage{Could not open file #1, ignoring it}%
    \else                       
        {
            \chardef\other=12
            \def\do##1{\catcode`##1=\other}%
            \dospecials
            \catcode`\ =10
            \epsffileoktrue         
            \epsfatendfalse     
            \loop               
                \read\epsffilein to \epsffileline
                \ifeof\epsffilein 
                \epsffileokfalse 
            \else                
                \expandafter\epsfaux\epsffileline:. \\%
            \fi
            \ifepsffileok
            \repeat
            \ifepsfbbfound
            \else
                \ifepsfverbose
                    \immediate\write16{No BoundingBox comment found in %
                                    file #1; using defaults}%
                \fi
            \fi
        }
        \closein\epsffilein
    \fi                         
    \epsfsetsize                
    \epsfstatus{#1}%
}%
\def\epsfclipoff{\def\epsfclipstring{\ifepsfdraft\space clip\fi}}%
\epsfclipoff 
\def\epsfspecial#1{%
     \epsftmp=10\epsfxsize
     \divide\epsftmp\pspoints
     \ifnum\epsfrsize=0\relax
       \includegraphics{\ifepsfdraft}%
     \else
       \epsfrsize=10\epsfysize
       \divide\epsfrsize\pspoints
       \includegraphics{\ifepsfdraft}%
     \fi
}%
\def\epsfframe#1%
{%
  \leavevmode                   
  \setbox0 = \hbox{#1}%
  \dimen0 = \wd0                                
  \advance \dimen0 by 2\epsfframemargin         
  \advance \dimen0 by 2\epsfframethickness      
  \vbox
  {%
    \hrule height \epsfframethickness depth 0pt
    \hbox to \dimen0
    {%
      \hss
      \vrule width \epsfframethickness
      \kern \epsfframemargin
      \vbox {\kern \epsfframemargin \box0 \kern \epsfframemargin }%
      \kern \epsfframemargin
      \vrule width \epsfframethickness
      \hss
    }
    \hrule height 0pt depth \epsfframethickness
  }
}%
\def\epsfsetgraph#1%
{%
   \leavevmode
   \hbox{
     \ifepsfframe\expandafter\epsfframe\fi
     {\vbox to\epsfysize
     {%
        \ifepsfshow
            \vfil
            \hbox to \epsfxsize{\epsfspecial{#1}\hfil}%
        \else
            \vfil
            \hbox to\epsfxsize{%
               \hss
               \ifepsfshowfilename
               {%
                  \epsfframemargin=3pt 
                  \epsfframe{{\tt #1}}%
               }%
               \fi
               \hss
            }%
            \vfil
        \fi
     }%
   }}%
   \global\epsfxsize=0pt
   \global\epsfysize=0pt
}%
\def\epsfsetsize
{%
   \epsfrsize=\epsfury\pspoints
   \advance\epsfrsize by-\epsflly\pspoints
   \epsftsize=\epsfurx\pspoints
   \advance\epsftsize by-\epsfllx\pspoints
   \epsfxsize=\epsfsize{\epsftsize}{\epsfrsize}%
   \ifnum \epsfxsize=0
      \ifnum \epsfysize=0
        \epsfxsize=\epsftsize
        \epsfysize=\epsfrsize
        \epsfrsize=0pt
      \else
        \epsftmp=\epsftsize \divide\epsftmp\epsfrsize
        \epsfxsize=\epsfysize \multiply\epsfxsize\epsftmp
        \multiply\epsftmp\epsfrsize \advance\epsftsize-\epsftmp
        \epsftmp=\epsfysize
        \loop \advance\epsftsize\epsftsize \divide\epsftmp 2
        \ifnum \epsftmp>0
           \ifnum \epsftsize<\epsfrsize
           \else
              \advance\epsftsize-\epsfrsize \advance\epsfxsize\epsftmp
           \fi
        \repeat
        \epsfrsize=0pt
      \fi
   \else
     \ifnum \epsfysize=0
       \epsftmp=\epsfrsize \divide\epsftmp\epsftsize
       \epsfysize=\epsfxsize \multiply\epsfysize\epsftmp
       \multiply\epsftmp\epsftsize \advance\epsfrsize-\epsftmp
       \epsftmp=\epsfxsize
       \loop \advance\epsfrsize\epsfrsize \divide\epsftmp 2
       \ifnum \epsftmp>0
          \ifnum \epsfrsize<\epsftsize
          \else
             \advance\epsfrsize-\epsftsize \advance\epsfysize\epsftmp
          \fi
       \repeat
       \epsfrsize=0pt
     \else
       \epsfrsize=\epsfysize
     \fi
   \fi
}%
\def\epsfstatus#1{
   \ifepsfverbose
     \immediate\write16{#1: BoundingBox:
                  llx = \epsfllx\space lly = \epsflly\space
                  urx = \epsfurx\space ury = \epsfury\space}%
     \immediate\write16{#1: scaled width = \the\epsfxsize\space
                  scaled height = \the\epsfysize}%
   \fi
}%
{\catcode`\%=12 \global\let\epsfpercent=
\global\def\epsfatend{(atend)}%
\long\def\epsfaux#1#2:#3\\%
{%
   \def\testit{#2}
   \ifx#1\epsfpercent           
       \ifx\testit\epsfbblit    
            \epsfgrab #3 . . . \\%
            \ifx\epsfllx\epsfatend 
                \global\epsfatendtrue
            \else               
                \ifepsfatend    
                \else           
                    \epsffileokfalse
                \fi
                \global\epsfbbfoundtrue
            \fi
       \fi
   \fi
}%
\def\epsfempty{}%
\def\epsfgrab #1 #2 #3 #4 #5\\{%
   \global\def\epsfllx{#1}\ifx\epsfllx\epsfempty
      \epsfgrab #2 #3 #4 #5 .\\\else
   \global\def\epsflly{#2}%
   \global\def\epsfurx{#3}\global\def\epsfury{#4}\fi
}%
\def\epsfsize#1#2{\epsfxsize}%